\documentclass[traditabstract,printer]{aa}
\usepackage{txfonts}
\usepackage{graphicx}
\usepackage{subfigure}
\usepackage{color}
\usepackage{natbib}

\begin{document}

\title{Detailed chemical abundance analysis of the thick disk star cluster Gaia 1
\thanks{This paper includes data gathered with the 2.5 meter du Pont Telescope located at Las Campanas Observatory, Chile.}}

% \subtitle{}

\author{
Andreas Koch\inst{1,2} 
  \and Terese T. Hansen\inst{3}
  \and Andrea Kunder\inst{4,5}
  }
  
\authorrunning{A. Koch, T.T. Hansen, \& A. Kunder}
\titlerunning{High-resolution abundance study of the thick disk cluster Gaia 1}
\offprints{A. Koch;  \email{a.koch1@lancaster.ac.uk}}
\institute{Department of Physics, Lancaster University, LA1 4YB, Lancaster, UK
\and Zentrum f\"ur Astronomie der Universit\"at Heidelberg, Astronomisches Recheninstitut, M\"onchhofstr. 12, 69120 Heidelberg, Germany
  \and Carnegie Observatories, 813 Santa Barbara St., Pasadena, CA 91101, USA
  \and Leibniz-Institut f\"ur Astrophysik Potsdam, An der Sternwarte 16, 14482 Potsdam, Germany   
  \and Saint Martin's University, Old Main, 5000 Abbey Way SE, Lacey, WA 98503, USA
   }
\date{Received: 26 June 2017 / Accepted: 11 September 2017}
\abstract{
Star clusters, particularly those objects in the disk-bulge-halo interface are as of yet poorly charted, albeit carrying important information about the 
formation and the structure of the Milky Way.  
Here, we present a detailed chemical abundance study of the recently discovered object Gaia~1. Photometry has previously 
suggested it as an intermediate-age, moderately metal-rich system, although the exact values for its age and metallicity remained ambiguous in the literature. 
We measured detailed chemical abundances of 14 elements in four red giant members, from high-resolution (R=25000) spectra that firmly establish Gaia~1 as an object 
associated with the thick disk. 
The resulting mean Fe abundance is $-0.62\pm$0.03(stat.)$\pm$0.10(sys.) dex, which is more metal-poor than indicated by  previous spectroscopy from the literature, but it is fully 
in line with values from  isochrone fitting. 
We find that Gaia~1 is moderately enhanced in the $\alpha$-elements, which allowed us to consolidate its membership with the thick disk via chemical tagging. 
The cluster's Fe-peak and neutron-capture elements  are similar to those found across the metal-rich disks, where the latter 
 indicate some level of $s$-process activity. 
No significant spread in iron nor in other heavy elements was detected, whereas we find evidence of light-element variations in  Na, Mg, and Al.
Nonetheless, the traditional  Na-O and Mg-Al  (anti-)correlations, typically seen in old globular clusters, are not seen in our data. 
This confirms that Gaia~1 is rather a massive and luminous open cluster than a low-mass globular cluster. 
Finally, orbital computations of the target stars bolster our chemical findings of Gaia 1's present-day membership with the thick disk, 
even though it remains unclear, which mechanisms put it in that place.  
 }
\maketitle 
%
%
%
%%%%%%%%%%%%%%%%%%%%%%%%%%%%%%%%%%%%%%%%%%%%%%%%%%%%%%%%%%%%%%%%%%%%%%%%%%%%%
%
%
\section{Introduction}
Star clusters in our Milky Way are excellent testbeds to study the formation, evolution, and structure of the underlying Galactic components. 
When turning to low Galactic latitudes, not only the, primarily old, globular clusters (GCs) are of interest, but young-to-intermediate-age open clusters (OCs) become the prime
tracer of the Galactic disks. Hence, a distinction between low-mass GCs and massive and luminous OCs is imperative. 
Likewise, establishing an association with either component can help with better charting the disk-bulge-halo interfaces in addition to studying 
their stellar content \citep[e.g.,][]{Ness2013,RecioBlanco2014,Koch2016,Koch2017ESO}. 

Gaia~1 is a star cluster that was recently discovered by \citet{Koposov2017} in the first Gaia data release \citep{GaiaDR1}, alongside with another system of lower mass.
Its observation and previous detections were seriously hampered by the  nearby bright star Sirius, which  emphasized the impressive discovery power of the {\em Gaia} mission. 
This object was first characterized as an intermediate-age (6.3 Gyr) and moderately metal-rich ($-$0.7 dex) system, based on isochrone fits to a comprehensive combination of {\em Gaia}, 2MASS 
\citep{Cutri2003}, WISE \citep{Wright2010}, and 
Pan-STARRS1 \citep{Chambers2016} photometry.
Thence,  this object was characterized by \citet{Koposov2017} as a star cluster, most likely of the globular confession.  
Further investigation of Gaia~1 found a metallicity higher by more than 0.5 dex, which challenged the previous age measurement and 
rather characterized it as a young (3 Gyr), metal-rich ($-0.1$ dex) object, possibly
of extragalactic origin given its orbit that leads it up to $\sim$1.7 kpc above the disk \citep{Simpson2017}. 
Subsequently, \citet{Mucciarelli2017} measured chemical abundances of six stars in Gaia~1, suggesting an equally high metallicity, but based on 
their abundance study, the suggestion of an extragalactic origin was revoked.
While a more metal-rich nature found by the latter authors conformed with the results by \citet{Simpson2017}, the evolutionary diagrams of both studies are very dissimilar and could not be explained 
by one simple isochrone fit. In particular, it was noted that ``the \citet{Simpson2017} stars do not define a red giant branch in the theoretical plane, suggesting that their parameters are not correct''
\citep[Fig.~1 of][]{Mucciarelli2017}. Such an inconsistency clearly emphasizes that a clear-cut chemical abundance scale is inevitable for 
fully characterising Gaia~1, and to further allow for tailored age determinations, even more so in the light of the seemingly well-determined orbital characteristics,  
Thus, this work focuses on a detailed chemical abundance analysis of four red giant members of Gaia~1, based on high-resolution spectroscopy, which we complement by 
an investigation of the orbital properties of this transition object. Combined with the red clump sample of \citet{Mucciarelli2017} and reaching down to the subgiant level \citep{Simpson2017}, 
stars in different evolutionary states in Gaia~1 
 are progressively being sampled.

This paper is organized as follows: In Sect.~2 we present our data acquisition and reduction, followed by 
a description of the ensuing chemical abundance analysis in Sect.~3. The results thereof are presented in Sect.~4.
In Sect.~5, we introduce our orbital computations for Gaia~1 before concluding on our results in Sect.~6.
\section{Observations}
Targets for our spectroscopy were selected from  the Two Micron All Sky Survey (2MASS) catalog \citep{Cutri2003} 
within an identical, fiducial spatial selection region as in  \citet[][see Fig.~1]{Koposov2017}. None of our targets overlap with the recent intermediate-resolution sample of 
\citet{Simpson2017}\footnote{ We 
note that the target source identification of the stars in the Tables~B.1 and B.2 of \citet{Simpson2017} is 
wrong in that there are no have cross-matches in the {\em Gaia} archive. 
The correct identifiers can be found in the csv tables provided by the authors 
in the preprint version of their paper (https://arxiv.org/e-print/1703.03823v1).}. 
\begin{figure}[htb]
\centering
\includegraphics[width=1\hsize]{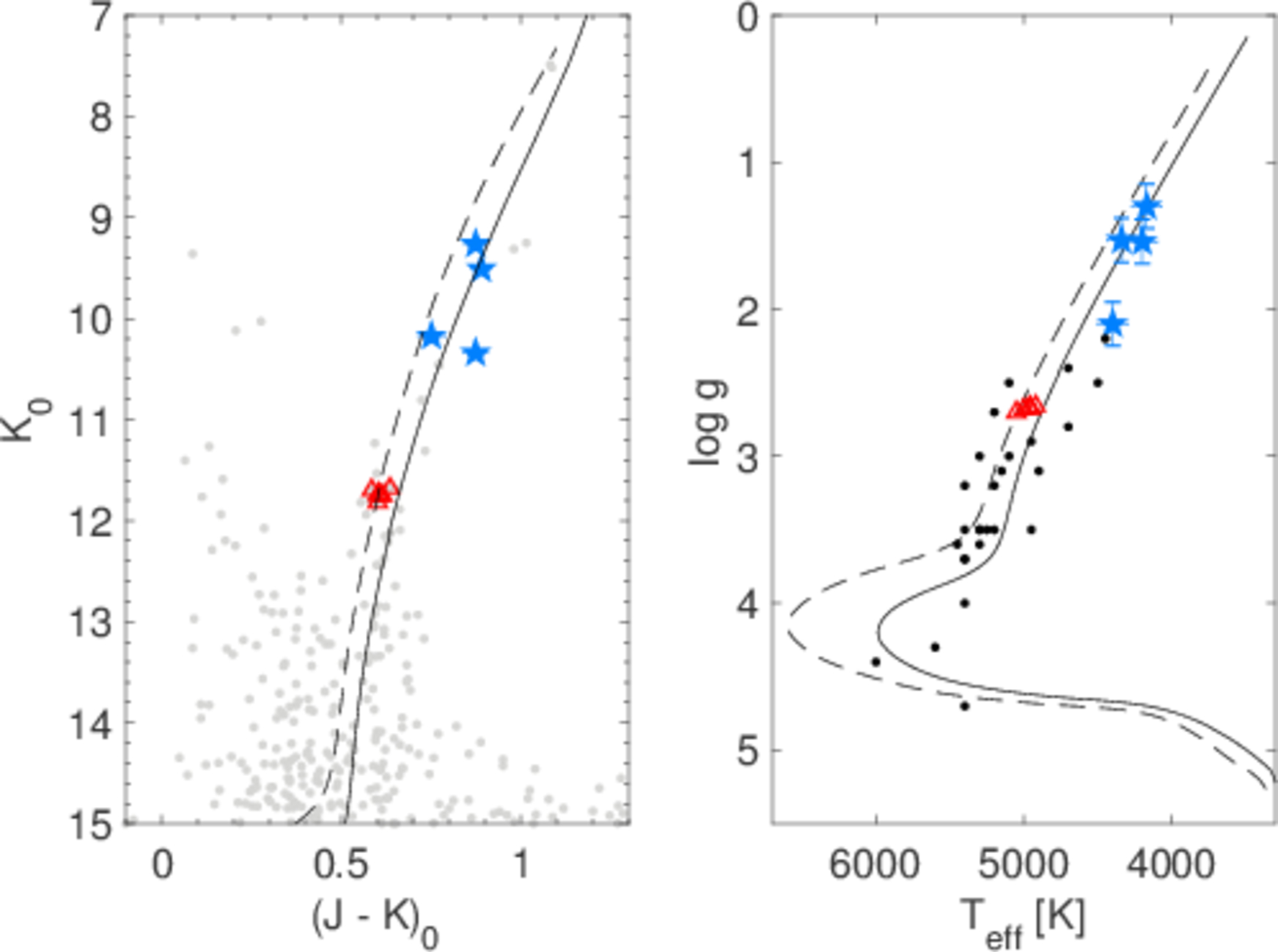}
\caption{Left panel: Colour magnitude diagram from 2MASS of stars within $\sim$2 half-light radii of Gaia 1 (gray points).  
Right panel: Hertzsprung-Russell diagram, constructed from our best spectroscopic stellar parameters. 
Our targets are highlighted in blue, red triangles are the He-clump targets of \citet{Mucciarelli2017}, and the sample of \citet{Simpson2017} is shown in 
black. Either panel also displays Dartmouth isochrones \citep{Dotter2008} using an age (6.3 Gyr) and metallicity ($-0.7$ dex), as 
suggested by \citet[][dashed line]{Koposov2017},  
and one with 12 Gyr and $-0.65$ dex (solid line).}
\end{figure}

Observations of four candidate members  were taken during four nights in March 2017
using the Echelle spectrograph at the 2.5-m du Pont telescope at the Las Campanas Observatory, Chile, with a  seeing of 0.7$\arcsec$--1.0$\arcsec$
throughout the nights.  
Our spectroscopic set-up included a 1.0$\arcsec$ slit with 2$\times$1 binning in spectral and spatial dimensions, resulting in a resolving power of R$\sim$25000. 
An observing log is given in Table~1.
\begin{table*}[htb]
\caption{Observing log and target properties}             
\centering          
\begin{tabular}{cccccccccc}     
\hline\hline       
  & $\alpha$ & $\delta$ & J & H & K & Date of & t$_{\rm exp}$ & SN\tablefootmark{a} & v$_{\rm HC}$ \\
\raisebox{1.5ex}[-1.5ex]{Star} & (J2000.0) & (J2000.0) & [mag] &  [mag] & [mag] & observation & [s]  & [px$^{-1}$] & [km\,s$^{-1}$] \\
\hline
1 & 06:45:49.77 & $-$16:44:46.47 & 10.550 &         9.700 &         9.440 & 06 Mar 2017 & 3$\times$1500 & 70/44/14 & 56.3$\pm$0.3 \\ 
2 & 06:45:57.64 & $-$16:40:11.00 & 10.816 &         9.965 &         9.687 & 07 Mar 2017 & 3$\times$1500 & 70/44/14 & 56.9$\pm$0.3 \\
3 & 06:45:53.81 & $-$16:45:32.21 & 11.337 & \llap{1}0.564 & \llap{1}0.350 & 08 Mar 2017 & 3$\times$1500 & 50/30/9 & 62.0$\pm$0.3 \\ 
4 & 06:45:55.93 & $-$16:43:46.98 & 11.628 & \llap{1}0.753 & \llap{1}0.518 & 09 Mar 2017 & 3$\times$1500 & 40/25/6 & 53.5$\pm$0.3 \\
\hline
\hline
\end{tabular} %
\tablefoot{
\tablefoottext{a}{Measured at 6550/5500/4500 ~\AA.}
}
\end{table*}
Despite the proximity to the nearby star Sirius, 
the location of our targets at 10$\arcmin$--12$\arcmin$  from this  bright  (V=$-1.5$ mag) object 
ensured that no stray light contaminated our spectra. In fact, no evidence for any flux from Sirius
is seen in our data, which would show up, e.g., in the form of blue-shifted Balmer lines (v$_{\rm Sir} =-5$ km\,s$^{-1}$ vs. $\sim$57 km\,s$^{-1}$ for the GC stars)
and/or an excess flux in the blue range, as Sirius' spectral type is A1V.

The data were reduced using the Carnegie Phython
Distribution\footnote{http://code.obs.carnegiescience.edu/} \citep{Kelson1998,Kelson2003,Kelson2000}. 
{ Subtraction of sky and scattered background light  was carefully treated by the software and aided by our use of a slit length of 4$\arcsec$ and no binning in spatial direction (see also Sect.~4.1.2).}
The final spectra  cover a full spectral range of 3340--8850~\AA. 
Typical  signal-to-noise (S/N) ratios reach 70 px$^{-1}$ in the peak of the order containing H$\alpha$ and decline towards 25  px$^{-1}$ in the bluer orders around 
4500~\AA~(Table 1). 
{ All four stars are radial velocity members of Gaia~1, and we will further discuss its kinematic properties in Sect.~5.1}.
\section{Abundance analysis}
Throughout the analysis we employed standard techniques as in our previous works \citep[e.g.,][]{Koch5897}. 
In brief, equivalent widths (EWs) were measured by fitting  Gaussian line profiles within IRAF's {\em splot} environment.
To this end, we used the line list from \citet{Ruchti2016} with updated log\,$gf$ values for Mg from \citet{PehlivanRhodin2017} and accounting for hyperfine splitting 
for odd-$Z$ elements. Due to the blending of features at the higher metallicity of the stars, we employed spectral synthesis for C, Ba, and Eu. 
The measurements of individual lines are summarized in Table~2.
\begin{table}[htbp]
\caption{Line list}
\centering          
\begin{tabular}{cccrrrrr}
\hline\hline       
& {$\lambda$} &  {E.P.} &  {} & \multicolumn{4}{c}{EW [m~\AA]} \\
\cline{5-8}
\raisebox{1.5ex}[-1.5ex]{Element} &  {[~\AA]} &  {[eV]}  &\raisebox{1.5ex}[-1.5ex]{log\,$gf$} 
&  {1} &  {2} &  {3} &  {4} \\
\hline
 O\,{\sc i} &  6300.30 &  0.000 &  $-$9.715 &  67 &  67 &  59 &  48 \\
 O\,{\sc i} &  6363.78 &  0.020 & $-$10.190 &  27 &  16 &  32 & \ldots \\
Na\,{\sc i} &  5682.63 &  2.102 &  $-$0.706 & 171 & 170 & 171 & 146 \\
Na\,{\sc i} &  5688.21 &  2.104 &  $-$0.404 & 192 & 178 & 174 & 164 \\
Na\,{\sc i} &  6154.23 &  2.102 &  $-$1.547 & 115 & 116 & 113 &  68 \\
Na\,{\sc i} &  6160.75 &  2.104 &  $-$1.246 & 125 & 121 & 135 & 100 \\
\hline
\hline
 \end{tabular}
\tablefoot{Table~2 is available in its entirety in electronic form via the CDS.}
\end{table}

All further analyses employed the 2014 version of the stellar abundance code MOOG \citep{Sneden1973}. 
As for the model atmospheres,  we interpolated Kurucz's grid\footnote{http://kurucz.harvard.edu/grids.html} 
of one-dimensional 72-layer, plane-parallel, line-blanketed models without convective overshoot, and assuming that 
local thermodynamic equilibrium (LTE) holds for all species. 
This model grid further incorporated the $\alpha$-enhanced opacity distributions, AODFNEW \citep{CastelliKurucz2004}. 
%http://wwwuser.oats.inaf.it/castelli/odfnew.html
%
Finally, all abundances were placed on the solar, photospheric scale of \citet{Asplund2009}.
\subsection{Stellar parameters}
For an initial estimate of the stellar temperature, we employed the calibrations for giants of  
 \citet{Alonso1999} using JHK 2MASS photometry, which, in turn, was converted to the TCS system needed for these calibrations.
 The  reddening maps of \citet{Schlafly2011} indicate an E(B$-$V)=0.49 mag with  negligible variation amongst the four targets. 
Finally, for the metallicity to enter the colour-T$_{\rm eff}$ relations we  estimated  the value from the isochrone fits of \citet{Koposov2017}, namely 
[Fe/H]=$-$0.7 dex. 
As a result, the temperatures from the (J$-$K) and (J$-$H) colour-indices agree to within 55 K, with the same order of 1$\sigma$ scatter. 
This compares to a typical error on the colour-temperatures of 70--140 K. 
Our photometric temperatures are listed in Table~3, alongside all other stellar parameters of our sample. 
\begin{table}[htbp]
\caption{Stellar parameters}
\centering          
\begin{tabular}{cccccc}
\hline\hline       
&   \multicolumn{3}{c}{T$_{\rm eff}$ [K]}  & log\,$g$ & $\xi$   \\
\cline{2-4}
\raisebox{1.5ex}[-1.5ex]{Star} & (J$-$H) & (J$-$K) & (spec.) & [dex] & [km\,s$^{-1}$] \\
\hline
1 & 4160 & 4081 & 4170 & 1.30 &  2.0   \\ 
2 & 4169 & 4047 & 4200 & 1.54 & 2.0   \\
3 & 4383 & 4368 & 4340 & 1.53 & 2.0 \\
4 & 4086 & 4083 & 4400 & 2.07 & 2.0  \\
\hline
\hline
 \end{tabular}
\end{table}

Next, T$_{\rm eff}$ was further refined by requiring  excitation equilibrium of neutral iron lines, where we culled
too strong and the weakest lines from consideration by requiring
$-5.5$\,$\la$\,$\log{\rm EW}/\lambda$\,$\la$\,$-4.5$. Furthermore, we restricted this part of the analysis to wavelengths redder than 5200~\AA~owing to the 
increasing noise at the bluer end.
The resulting spectroscopic temperatures could be typically determined to within 100 K (based on the line-to-line scatter in 
comparison with the change of slopes) and 
are in excellent agreement with the photometric ones, with the exception of star \#4, which shows a colder photometric colour.
For the remainder of the analysis we used the spectroscopic values as the final T$_{\rm eff}$ in the atmospheres. 

In the next step, photometric gravities were determined by adopting the above temperatures, a distance to Gaia 1 of 4.6 kpc \citep{Koposov2017} and 
assuming a typical mass of the stars on the RGB of 1.0 M$_{\odot}$, as indicated by an intermediate-age, metal-rich Dartmouth isochrone \citep{Dotter2008}
based on the stellar population identified by \citet{Koposov2017}. By enforcing ionization equilibrium between neutral and ionized iron lines, we were also able to estimate
spectroscopic surface gravities, which are in good agreement with the photometric ones. The implications of our refined log\,$g$ will be discussed further in Sect.~5.

Finally, the microturbulence was fixed by removing the trend of the abundance from neutral iron lines with reduced width, $\log{\rm EW}/\lambda$. 
As before, weak lines were removed from this exercise using the same cuts as mentioned above, since they are increasingly prone to noise.
Likewise, strong lines  
were deemed unsuitable due to their saturation on the flat part of the curve of growth \citep{Magain1984,Hanke2017}. 
Overall, the resulting abundance trend was not very sensitive to the choice of microturbulence, placing all our stars around values of $2.0\pm0.2$ km\,s$^{-1}$. 
\subsection{Abundance errors}
Also for the abundance errors we followed a standard approach. To this end,  statistical errors were adopted via the line-to-line scatter from multiple line measurements, which we list as $\sigma$ in Table~4, together with the number of lines, $N$, that were measured. 
The systematic errors, in turn,  were quantified via variations of the stellar atmosphere parameters. Thus we computed new abundances after varying each parameter independently about their 
uncertainty, i.e., T$_{\rm eff}\pm$100 K; log\,$g\pm$0.15 dex; $\xi\pm0.2$ km\,s$^{-1}$; [M/H]$\pm$0.1 dex, and switching from the $\alpha$-enhanced opacity distributions to the Solar-scaled values. 
For the latter we adapted half the abundance change as the implied error, mimicking an uncertainty in the atmospheres' [$\alpha$/Fe] ratio of $\pm$0.2 dex. 
All respective changes and 
the overall result, obtained by adding the contributions in quadrature,  are summarized in Table~5, exemplary for star \#1. 
\begin{table*}[htb]
\caption{Abundance results}             
\centering          
\begin{tabular}{crcrcrcrcrcrcrcrccc}     
\hline\hline       
  &   &  1  &   &   &   &  2 &   &   &   
&  3  &   &   &   &  4  &  & & \multicolumn{2}{c}{Gaia~1}   \\
\cline{2-4}\cline{6-8}\cline{10-12}\cline{14-16}\cline{18-19}
\raisebox{1.5ex}[-1.5ex]{Element\tablefootmark{a}} &  [X/Fe]  &  $\sigma$ &  N  &   &  
 [X/Fe] &  $\sigma$ &  N  &   &   [X/Fe] &  $\sigma$ &  N  &   &  [X/Fe]&   $\sigma$ &  N & & [X/Fe]\tablefootmark{c} & $\sigma$\tablefootmark{c}\\
 \hline
Fe\,{\sc i}        	       & $-$0.67 &   0.21 & 97     && $-$0.55 &   0.22 &     89 && $-$0.67 &   0.24 &     94 & & $-$0.60 &   0.28 &    88  &&  \llap{$-$}0.62$\pm$0.03 & 0.05$\pm$0.02 \\
Fe\,{\sc ii}       	       & $-$0.66 &   0.33 &  9     && $-$0.54 &   0.27 &      9 && $-$0.66 &   0.22 &      9 & & $-$0.61 &   0.22 & 	7  &&   \llap{$-$}0.61$\pm$0.08 & 0.00$\pm$0.08 \\
C\,{\sc i}\tablefootmark{b}    & $-$0.21 &   0.40 & \ldots && $-$0.53 &   0.40 & \ldots && $-$0.38 &   0.40 & \ldots & & $-$0.85 &   0.40 & \ldots &&   \llap{$-$}0.49$\pm$0.20 & 0.00$\pm$0.25 \\
O\,{\sc i}       	       &    0.36 &   0.03 &  2     &&    0.25 &   0.15 &      2 &&    0.49 &   0.10 &      2 & &    0.50 & \ldots & 	1  &&   $0.41\pm$0.04 & 0.06$\pm$0.04 \\
Na\,{\sc i}\rlap{$^{\rm LTE}$} &    0.53 &   0.07 &  4     &&    0.36 &   0.14 &      4 &&    0.64 &   0.18 &      4 & &    0.24 &   0.06 & 	4  &&   $0.43\pm$0.08 & 0.14$\pm$0.06 \\
Na\,{\sc i}\rlap{$^{\rm NLTE}$}&    0.38 &   0.08 &  4     &&    0.21 &   0.16 &      4 &&    0.48 &   0.16 &      4 & &    0.10 &   0.06 & 	4  &&   $0.29\pm$0.07 & 0.14$\pm$0.06 \\
Mg\,{\sc i}       	       &    0.41 &   0.20 &  8     &&    0.15 &   0.17 &      7 &&    0.29 &   0.23 &      7 & &    0.14 &   0.19 & 	7  &&   $0.24\pm$0.06 & 0.09$\pm$0.05 \\
Al\,{\sc i}       	       &    0.64 & \ldots &  1     &&    0.46 & \ldots &      1 &&    0.31 & \ldots &      1 & &    0.46 & \ldots & 	1  &&   $0.47\pm$0.06 & 0.11$\pm$0.05 \\
Si\,{\sc i}       	       &    0.36 &   0.27 & 17     &&    0.31 &   0.36 &     17 &&    0.35 &   0.30 &     15 & &    0.26 &   0.27 &    16  &&   $0.32\pm$0.04 & 0.00$\pm$0.05 \\
Ca\,{\sc i}       	       &    0.22 &   0.29 & 14     &&    0.04 &   0.29 &     13 &&    0.25 &   0.35 &     13 & &    0.19 &   0.34 &    14  &&   $0.17\pm$0.04 & 0.00$\pm$0.09 \\
Sc\,{\sc ii}      	       &    0.05 &   0.23 &  4     &&    0.17 &   0.13 &      4 &&    0.13 &   0.29 &      3 & &    0.01 &   0.07 & 	3  &&   $0.07\pm$0.05 & 0.05$\pm$0.05 \\
Ti\,{\sc i}       	       &    0.35 &   0.36 & 28     &&    0.17 &   0.20 &     21 &&    0.11 &   0.14 &     23 & &    0.22 &   0.24 &    20  &&   $0.20\pm$0.04 & 0.07$\pm$0.04 \\
Ti\,{\sc ii}      	       &    0.14 &   0.23 &  5     && $-$0.07 &   0.18 &      4 && $-$0.09 &   0.09 &      4 & & $-$0.16 &   0.06 & 	3  &&   \llap{$-$}0.09$\pm$0.05 & 0.05$\pm$0.06 \\
Cr\,{\sc i}       	       &    0.16 &   0.26 & 10     && $-$0.14 &   0.27 &     10 && $-$0.08 &   0.19 &      9 & &    0.04 &   0.29 &    10  &&   \llap{$-$}0.01$\pm$0.06 & 0.08$\pm$0.06 \\
Co\,{\sc i}                    &    0.25 &   0.29 &  7     &&	 0.04 &   0.21 &      7 &&    0.13 &   0.35 &      6 & &    0.30 &   0.27 & 	4  &&   $0.15\pm$0.06 & 0.04$\pm$0.10 \\
Ni\,{\sc i}       	       &    0.08 &   0.35 & 36     && $-$0.02 &   0.35 &     37 &&    0.02 &   0.31 &     36 & & $-$0.01 &   0.30 &    32  &&   $0.02\pm$0.05 & 0.00$\pm$0.05 \\
Ba\,{\sc ii}                   & $-$0.18 &   0.07 &  3     && $-$0.14 &   0.08 &      3 && $-$0.22 &   0.07 &      3 & & $-$0.04 &   0.10 & 	3  &&   \llap{$-$}0.18$\pm$0.07 & 0.00$\pm$0.07 \\
Eu\,{\sc ii}	  	       &    0.11 &   0.05 &  1     &&    0.05 &   0.05 &      1 &&    0.12 &   0.10 &      0 & &    0.28 &   0.15 & 	1  &&   $0.11\pm$0.05 & 0.00$\pm$0.05 \\
 \hline 				    
 \hline
 \end{tabular}
\tablefoot{
\tablefoottext{a}{Ionised  species are given relative to Fe\,{\sc ii}. 
Abundance ratios are listed relative to iron, except for Fe\,{\sc i}  and Fe\,{\sc ii} (relative to H).}
\tablefoottext{b}{From the CH G-band.}
\tablefoottext{c}{Mean and 1$\sigma$-dispersion of the entire cluster based on an error-weighted maximum likelihood approach.}
}
\end{table*}
\begin{table*}[htb]
\caption{Systematic errors for Star 1}             
\centering          
\begin{tabular}{ccccccc}     
\hline\hline       
  & T$_{\rm eff}$ & log\,$g$ & $\xi$ & [M/H] &  &  \\
 \raisebox{1.5ex}[-1.5ex]{Element} & $\pm100$ K & $\pm0.15$ dex & $\pm0.2$ km\,s$^{-1}$ & $\pm$0.1 dex & \raisebox{1.5ex}[-1.5ex]{ODF} & \raisebox{1.5ex}[-1.5ex]{$\sigma_{\rm sys, tot}$} \\
 \hline
Fe\,{\sc i}   &   $<$0.01 & $\pm$0.03 & $\pm$0.09  & $\pm$0.02 & $-$0.07 & 0.10 \\
Fe\,{\sc ii}  & $\mp$0.16 & $\pm$0.07 & $\pm$0.04  & $\pm$0.03 & $-$0.16 & 0.20 \\
C\,{\sc i}    & $\mp$0.19 & $\mp$0.08 & $\pm$0.14  &   $<$0.01 & $-$0.08 & 0.26\\
O\,{\sc i}    & $\pm$0.02 & $\pm$0.06 &   $<$0.01  & $\pm$0.04 & $-$0.12 & 0.10 \\
Na\,{\sc i}   & $\pm$0.09 & $\mp$0.01 & $\pm$0.10  &   $<$0.01 & $-$0.03 & 0.13 \\
Mg\,{\sc i}   & $\pm$0.03 &   $<$0.01 & $\pm$0.05  & $\pm$0.02 & $-$0.06 & 0.07 \\
Al\,{\sc i}   & $\pm$0.05 &   $<$0.01 & $\pm$0.03  &   $<$0.01 & $-$0.02 & 0.06 \\
Si\,{\sc i}   & $\mp$0.09 & $\pm$0.03 & $\pm$0.03  & $\pm$0.02 & $-$0.09 & 0.11 \\
Ca\,{\sc i}   & $\pm$0.11 & $\mp$0.01 & $\pm$0.14  &   $<$0.01 & $-$0.05 & 0.18 \\
Sc\,{\sc ii}  & $\mp$0.03 & $\pm$0.06 & $\pm$0.06  & $\pm$0.03 & $-$0.13 & 0.11 \\
Ti\,{\sc i}   & $\pm$0.16 &   $<$0.01 & $\pm$0.14  &   $<$0.01 & $-$0.06 & 0.22 \\
Ti\,{\sc ii}  & $\mp$0.03 & $\pm$0.06 & $\pm$0.13  & $\pm$0.03 & $-$0.13 & 0.16 \\
Cr\,{\sc i}   & $\pm$0.13 &   $<$0.01 & $\pm$0.09  &   $<$0.01 & $-$0.05 & 0.16 \\
Co\,{\sc i}   & $\pm$0.08 &   $<$0.01 & $\pm$0.13  & $\pm$0.02 & $-$0.07 & 0.16 \\
Ni\,{\sc i}   &   $<$0.01 & $\pm$0.04 & $\pm$0.10  & $\pm$0.02 & $-$0.09 & 0.11 \\
Ba\,{\sc ii}  & $\pm$0.02 & $\pm$0.04 & $\pm$0.11  & $\pm$0.04 & $-$0.16 & 0.15 \\
Eu\,{\sc ii}  & $\mp$0.02 & $\pm$0.06 & $\pm$0.04  & $\pm$0.03 & $-$0.12 & 0.10 \\
 \hline
 \hline
 \end{tabular}
\end{table*}
\section{Abundance results}
All abundance ratios are summarized in Table~4.
\subsection{Iron}
From our abundance analysis of the four Gaia~1 stars we found a mean iron abundance of $-0.62\pm$0.03(stat.)$\pm$0.10(sys.) dex. Here, ionization equilibrium has been achieved per construction in that we forced the Fe\,{\sc i} and Fe\,{\sc ii} abundances to agree in order to determine spectroscopic gravities. 
The 1$\sigma$ dispersion of our [Fe/H] is 0.05$\pm$0.02 dex, a low value fully in line with other stellar systems of similarly low mass (2.2$\times 10^4$ M$_{\odot}$;  $M_V\sim -5$ mag;  \citealt{Carretta2009Fe,Pancino2010,Koch2012}).
The moderately metal-rich value we found is in excellent agreement with the isochrone fitting of \citet{Koposov2017}, who adopted an intermediate age of the GC of 6.3 Gyr. 
{ However, the latter should be taken with caution, given the uncertain reddening 
of the cluster \citep{Mucciarelli2017} and the fact that the metallicity sensitivity of the RGB relies entirely on its colour. 
In addition, the available ground-based photometry does not  reach below the metallicity- and age-sensitive turn-off, 
which would allow for an understanding of its CMD morphology, and it is affected badly 
by the presence of Sirius. Future {\em Gaia} releases will provide  a better 
grip on the CMD of this cluster}. 
\subsubsection{ Comparison with medium-resolution literature results}
In stark contrast, \citet{Simpson2017} found a much more metal-rich value of $-0.13\pm0.13$ dex %(and a 1$\sigma$ spread of dex), 
using two approaches on   27 red clump member candidates.
These comprised an analysis at similar spectral resolution to the present work, but employing a smaller wavelength range, complemented by calcium triplet  (CaT) spectroscopy at low resolution
(R$\sim$10000). 
The cause for these very high metallicities from the high-resolution spectra lies beyond the scope of our analysis, but reasons could be related to S/N, which is listed down to negative values 
in \citet[][their Table~B.1]{Simpson2017}. 
We can furthermore conjecture 
on the employed CaT calibrations in that study (valid up to [Fe/H]=0.5, but dependent on stellar luminosity) or residual foreground contamination, for which the CaT calibrations cannot be applied. 
Furthermore, the majority of AAOMega targets in that study are stars on the lower RGB and subgiants, 
for which the CaT calibrations have only have limited applicability. 
Combined with the ambiguous target identification mentioned above (Sect.~2) and \citet{Mucciarelli2017}'s
suggestion of wrong stellar parameters in that study, these results are difficult to reconcile with our measurements. 
\subsubsection{ Comparison with high-resolution literature results}
Recently, \citet{Mucciarelli2017} performed an abundance analysis of six He-clump stars in Gaia~1 at a data quality comparable to our study, albeit at higher resolution of 36000.
Their main conclusion was that Gaia~1 is a disk object at Solar metallicity with Solar-scaled abundance ratios. 
While the metal-rich nature found by these authors conformed with the results by \citet{Simpson2017}, the HRD of both studies are very dissimilar and could not be explained 
by a simple isochrone fit. In particular, it was noted that ``the \citet{Simpson2017} stars do not define an RGB in the theoretical plane, suggesting that their parameters are not correct''
\citep[Fig.~1 of][and Fig.~1 in our present work]{Mucciarelli2017}. Such an inconsistency clearly emphasizes that age and metallicity derivations are vital, and we suggest that 
a lower metallicity as found in our analysis would be able to resolve the discrepancy with the literature values. 

\vspace{2ex}
\noindent
{\em  Instrumental effects}

\vspace{1ex}
\noindent
{ In order to test, whether the discrepant metallicity scales are an instrumental artefact,  we thoroughly checked the background subtraction performance of our data.
The typical sky level is $\sim$12--14\% of the object flux in the center of the orders around 6000 \AA~so that poor sky subtraction would have rendered our EWs lower by a
factor of up to 0.85, leading to even lower measured metallicities. 
Conversely, we would need to increase our EWs by 44\% to reach the Solar iron abundance measured by \citet{Mucciarelli2017}, whereas
both their and our studies reached a much higher accuracy.  
Due to our used slit length of 4$\arcsec$ and no binning in the spatial direction,  the subtraction of  sky and scattered light 
(note, e.g., a possible  contribution from a 79\% moon at 36$\degr$) was fully removed by our reduction procedures \citep{Kelson1998,Kelson2003,Kelson2000}\footnote{ Since 
our observations were taken during the same nights as those of \citet{Mucciarelli2017}, this would have affected both data sets in an identical manner.}.
}
\vspace{1ex}
\noindent
{\em  Mass loss}

\vspace{1ex}
\noindent
{  Asserting that the two abundance scales established by \citet{Mucciarelli2017} and ourselves are real, we need to consider the option that 
Gaia 1 hosts two stellar populations -- one metal-rich ($\sim$Solar) and one metal-poor ($\sim -0.6$ dex). Since \citet{Mucciarelli2017} 
focused on the He-burning red clump, which is, by definition, young and metal-rich, their study would have missed the presence 
of the metal-poor population that showed up in our RGB sample. 

This scenario, however, is uncomfortable given the system's low mass.
 Multiple populations of such distinction are mainly found in massive systems \citep[e.g.,][]{Lee2009,Johnson2010,Marino2015}, also towards the Galactic plane \citep[e.g.,][]{Mauro2012,Origlia2013}, 
 which, in light of Gaia~1's rather low mass would imply that it has lost a great deal of mass. 
This was deemed unrealistic by \citet{Mucciarelli2017} given the cluster's purported young age of 3--6 Gyr, while 
an older age cannot be ruled out by isochrones of a lower metallicity and higher age (Fig.~1). 
Moreover, the lack of a Na-O anti-correlation (Sect.~4.3) appears to argue against a more massive origin of Gaia~1. 
}
\subsection{Light elements: Li,C}
While we were able to obtain upper limits for the lithium abundance from synthesis of the 6707~\AA~resonance line, the low values  of typically 
A(Li)$<$$-0.3$  reflect the expected depletion seen in evolved giants \citep{Spite1982,Lind2009}. We do not consider this fragile element any further. 

Carbon abundances where derived from synthesis of the CH band at 4300 \AA, 
using the line list from \citet{Masseron2014}. We find carbon abundances that
are much lower than typical Galactic stars from the literature \citep[e.g.,][]{Nissen2014}, but those samples are mostly based on dwarfs, while our
evolved giants (log(L/L$_{\odot}$)$\sim$2--2.6) need  to be corrected for
evolutionary status \citep{Placco2014}. Such corrections for metal-poor stars
with similar T$_{\rm eff}$ and $\log g$ are as large as 0.7~dex, which would place our stars with
solar (or slightly enhanced) carbon abundances.
\subsection{Light element variations: O, Na, Mg, Al}
For determining O-abundances, we made use of the, generally strong, [O\,{\sc i}] 6300 and 6363~\AA~lines. 
Na abundances were derived from four transitions, neglecting the strong NaD lines. 
Each measurement was corrected for NLTE effects using the calculations of \citet{Lind2011} in order to aid a fair comparison with the literature.
Finally, our Al-abundance is based on the EW of the moderately strong (50--80 m\AA) 6698~\AA~line.

Highlighted in Fig.~2 are the probability distributions for the mean value (here taken relative to the nominal sample mean, $<$[X/Fe]$>$, to emphasize 
the actual abundance spreads) and the 1$\sigma$ dispersion over our entire sample, in each measured element, accounting for the individual measurement errors. 
The mean and dispersions for the entire cluster are also listed in Table~4. 
\begin{figure}[htb]
\centering
\includegraphics[width=1\hsize]{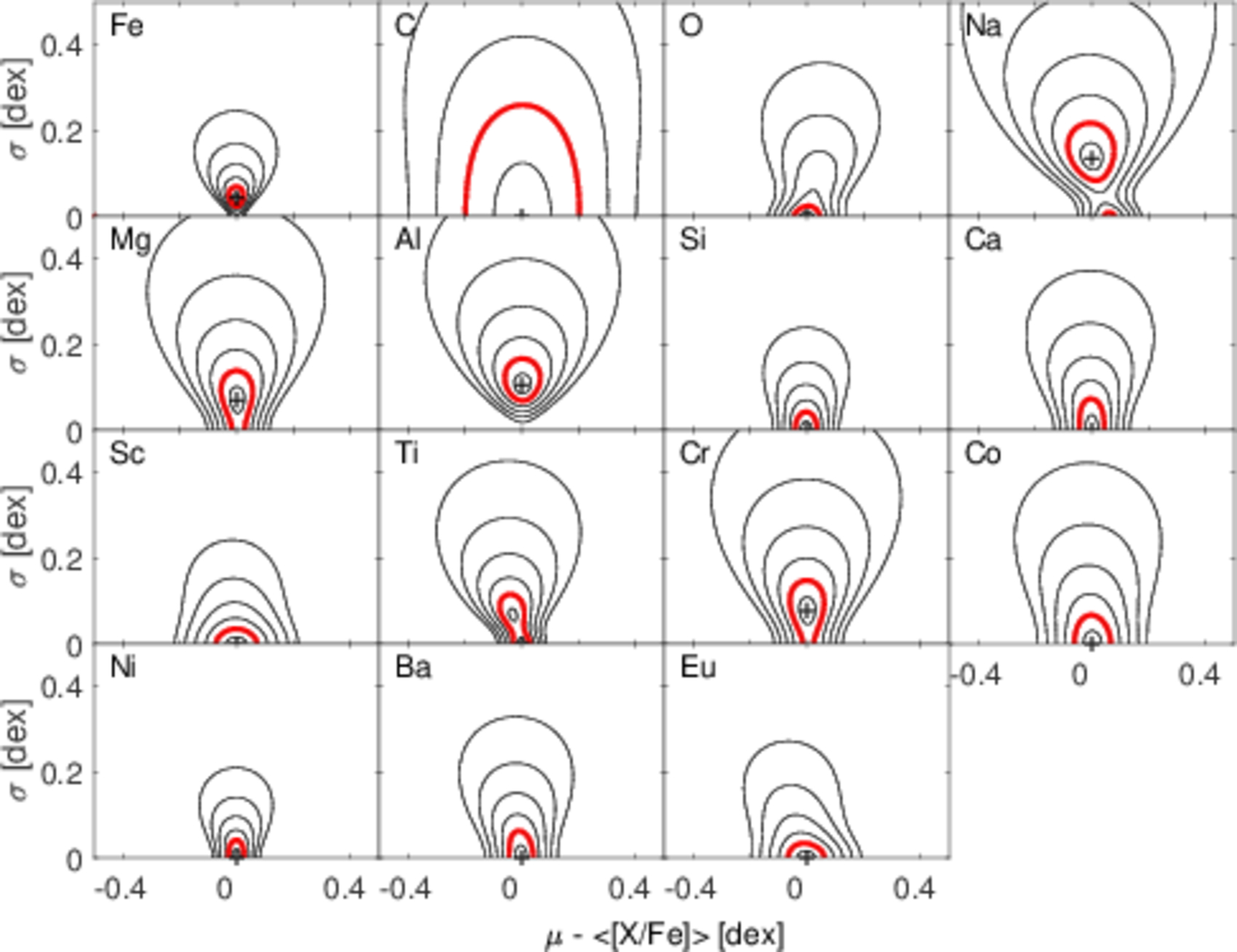}
\caption{Probability distributions of the mean values ($\mu$) and dispersions ($\sigma$) in the abundance ratios. Here, the x-axis has been normalized to zero at the observed 
mean of each element's [X/Fe] to emphasize the abundance spreads. 
Contours are placed at 0.5--3$\sigma$ in steps of 0.5, where the red curve highlights the 1$\sigma$ contour. Note the 
significant spreads in the light elements Na, Mg, and Al.}
\end{figure}

As expected for a chemically homogeneous system such as a star cluster, most elements have dispersions consistent with zero, i.e.,  any spread is driven by the measurement errors. 
The clear exceptions are those light elements that commonly partake in the proton-capture reactions in massive stars, i.e.,  Mg, Na, Al
 with a significant non-zero spread.

In Fig.~3 we further explore (anti-)correlations between these light elements that are found in the majority of Galactic GCs
as a consequence of hot proton-burning reactions in a first generation of massive stars. These trends, however, are 
 are markedly lacking in Galactic field and OC stars, {  and in dwarf galaxies} \citep{Geisler2007}. 
\begin{figure}[htb]
\centering
\includegraphics[width=1\hsize]{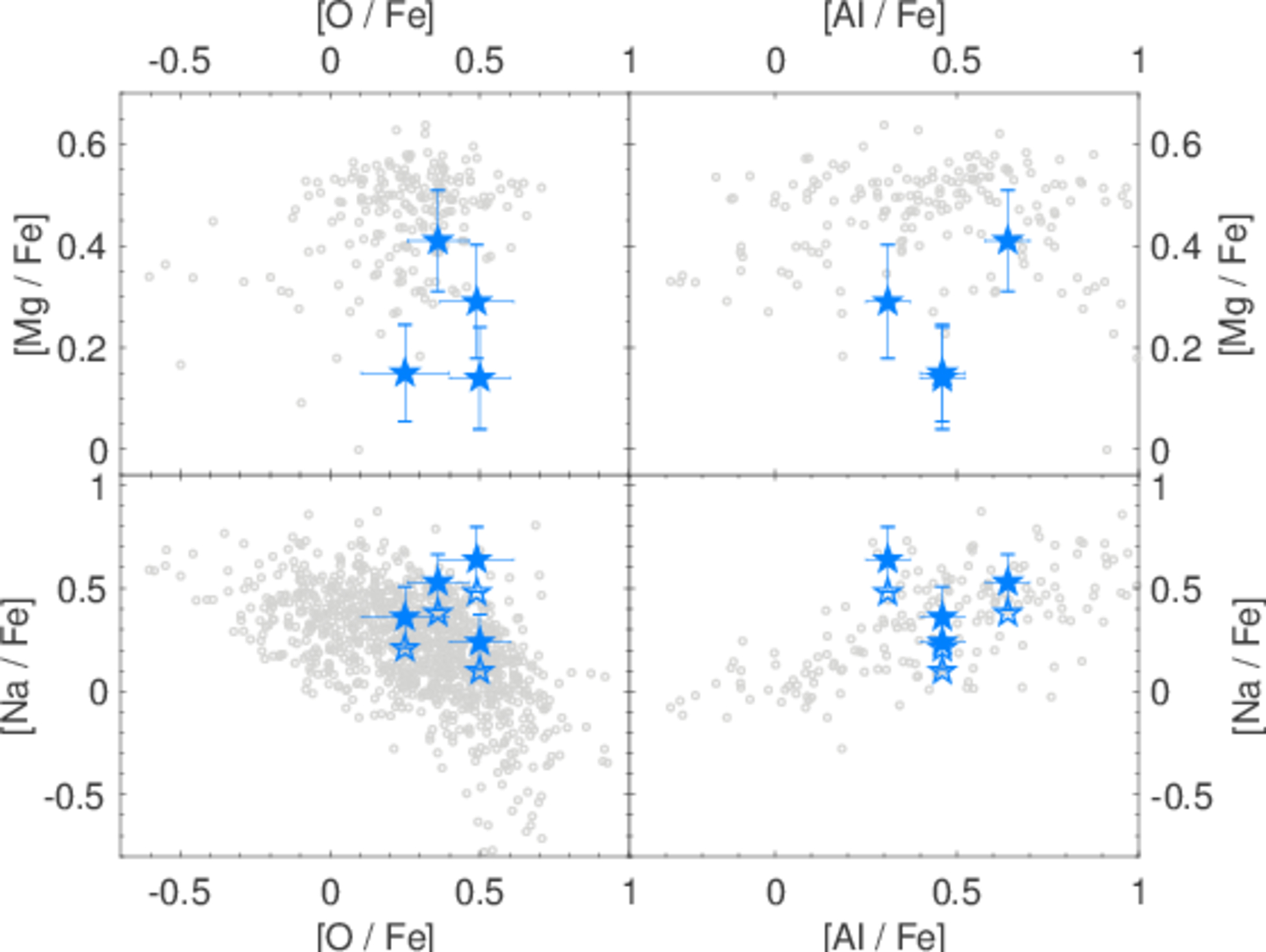}
\caption{
Juxtaposition of various light element (anti-) correlations. 
GC stars from \citet[]{Carretta2009NaO} are indicated as gray circles, while 
our measurements are shown in blue. Here, open symbols are the NLTE-corrected values (Na) to aid a fair comparison with the literature.
The correlations in Gaia~1 are weak, but a large spread in all elements is clearly visible.} 
\end{figure}
As Fig.~3 implies, there is no pronounced anti-correlation in Gaia~1, although both Na and O show a broad spread. 
Thus we see, chemically,  no evidence for any multiple populations in our sample alone, while arguably this can also be driven by the low number of targets.
Likewise, no significant Mg-Al correlation is seen in our data, although this trend is generally much weaker in the GCs than the more prominent Na-O relation, 
to the point, where it breaks down for the lowest-mass systems and/or at the metal-rich tail of the GC population  \citep[e.g.,][]{Pancino2017}.

At a present stellar mass of a few 10$^4$ M$_{\odot}$, Gaia 1 lies at the boundary between a very low-mass GC and a very massive OC \citep[e.g.,][]{Bragaglia2012}, 
which corroborates previous findings of a threshold cluster mass, below which light element correlations are not detected \citep{Carretta2010}.
In this context,  \citet{Pancino2010} note that no evidence of light element variations in excess of  $\la$0.2 dex is seen in their sample of OCs.
Interestingly, the apparent lack of a Na-O anti-correlation, despite the presence of abundance spreads in Na and Al, 
in a sample of seven stars in the bulge GC NGC 6440 \citep{Munoz2017}, 
closely resembles our finding in Gaia~1. Although, at 5.7$\times$10$^5$ M$_{\odot}$, the former is more massive than Gaia~1 by a factor of several tens, this 
bolsters the suggestion of  \citet{Munoz2017} that some bulge clusters underwent a  chemical evolution different from the rest of the Milky Way GC population. 
{ If Gaia~1 had developed from a once more massive system, which is feasible given the potential bimodal metallicity distribution, 
its progenitor could have resembled the massive NGC 6440 with its lack of a Na-O relation. }
\subsection{$\alpha$-elements: Mg, Si, Ca, Ti}
In Figs.~4--8 we overplot our measurements on literature data for various Galactic components.
Specifically, these contain 
disk stars from \citet{Koch2002,Reddy2003,Reddy2006,Bensby2014,Battistini2016},
and bulge samples from \citet{Johnson2012,Bensby2013,Johnson2014,Vanderswaelmen2016}.

Gaia~1 is moderately $\alpha$-enhanced to 0.2--0.3 dex, depending on the exact element considered, thereby reflecting the different contributing nucleosynthetic channels. 
Its location in abundance space is shown in Fig.~4 in comparison with the Galactic disks and bulge. 
The [Ca/Fe] abundance ratios are systematically lower than the remaining elements, where the largest values of $\sim$0.3 dex are reached for Si. 
The elevated [$\alpha$/Fe] ratios are broadly consistent with a significant contribution from  SNe II. 
\begin{figure}[htb]
\centering
\includegraphics[width=1\hsize]{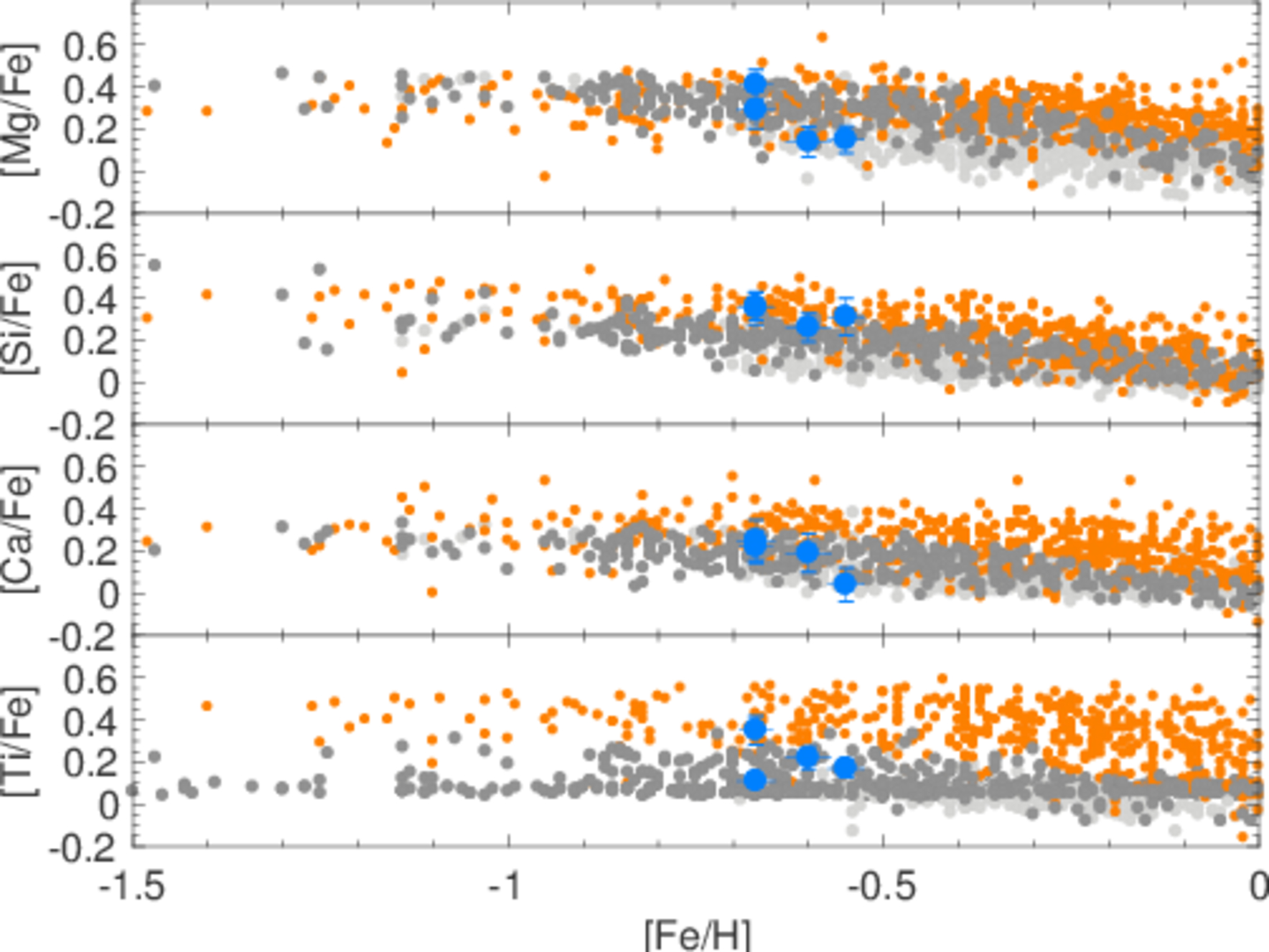}
\includegraphics[width=1\hsize]{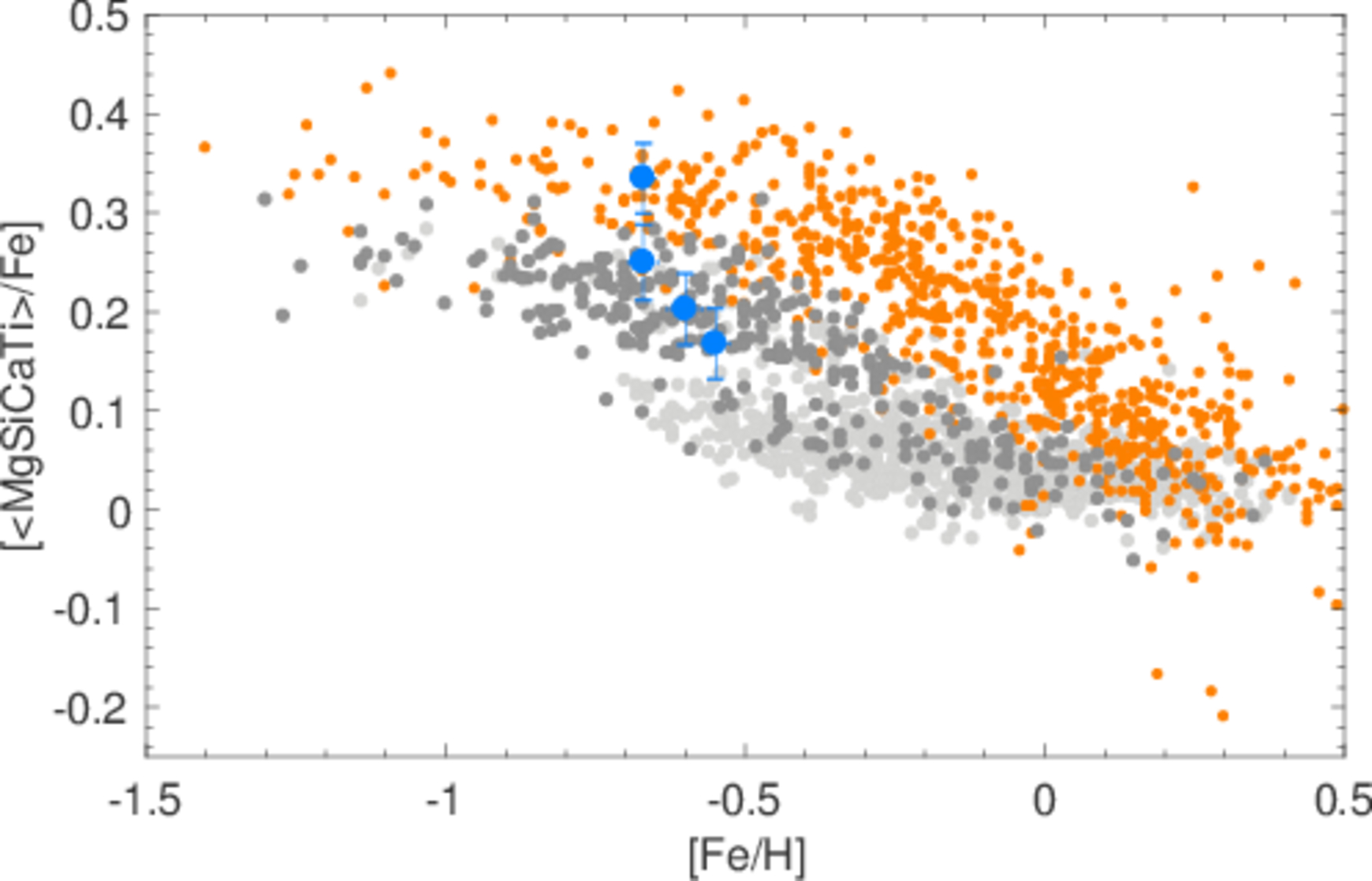}
\caption{Abundance ratios of the $\alpha$-elements. Our stars are shown as blue symbols. Orange dots denote 
bulge stars from \citet{Gonzalez2011,Bensby2013,Johnson2014}, while disk stars are depicted as gray points \citep{Reddy2003,Reddy2006,Bensby2014}.
Here, we specifically distinguish between thin (light gray) and thick disk stars (dark gray).
The bottom panel shows the straight average of the Mg, Si, Ca, and Ti abundances as a proxy for $\alpha$-enhancement. 
In all panels, the axis have been truncated to highlight the abundance region of our targets.}
\end{figure}

If we simplistically average over all four elements (bottom panel of Fig.~4), disregarding the difference in their hydrostatic vs. explosive nature \citep[cf.][]{Koch2008,McWilliam2016}, 
we find a mean $\alpha$-enhancement of [$<$Mg,Si,Ca,Ti$>$/Fe]=0.24$\pm$0.03 dex with a 1$\sigma$ spread of 0.05$\pm$0.03 dex. 
Considering that the latter is predominantly driven by a significant spread in Mg as a consequence of the proton-capture reactions within the Mg-Al cycle, 
this provides no evidence of any spread in the majority of $\alpha$-elements. 

Fig.~4 (bottom) shows a clear separation of the Galactic components. Here, the literature values of \citet{Reddy2006} and \citet{Bensby2014} allowed for a unique separation into the 
thin and thick disks purely based on stellar kinematics \citep[see also][and references therein]{RojasArriagada2017,Feltzing2013}. 
This comparison leaves little doubt about Gaia~1 being associated with the thick disk in terms of its moderate enhancement between the values
in bulge stars (near the halo plateau of $\sim$0.4 dex) and the thin disk abundances, only slightly in excess of Solar values. 
Here we note that this kind of comparison does not account for spatial variations in the literature samples, ignoring the effects 
of  any possible metallicity gradients  \citep[e.g.,][]{Janes1979,Chen2003,Magrini2009,Cunha2016,Reddy2016}.

Likewise, we do not detect any significant spreads in Si or Ca, 
as has been suggested for some bulge and (metal-poor) halo GCs 
as a consequence of leakage of protons from the Mg-Al chain in massive stars \citep{Carretta2009NaO,Hanke2017,Munoz2017}. 
\subsection{Fe-peak elements: Sc, V, Cr, Co, Ni, Zn}
Fig.~5 shows our measurements for the Fe-peak elements Sc, Cr, Co, and Ni along with the literature data for the bulge and thin and thick disk.
None of these has been corrected for effects of NLTE, as none of the reference sets has been. Corrections for Cr are expected to be negligible in our stars \citep{Bergemann2010Cr}, while 
 Co-abundances have NLTE corrections as high as 0.3 dex in thick disk stars, albeit at higher T$_{\rm eff}$ ($\sim$6000 K) than in our sample, or even up to 0.6 dex for cool, metal-poor halo stars  \citep{Bergemann2010Co}. 
\begin{figure}[htb]
\centering
\includegraphics[width=1\hsize]{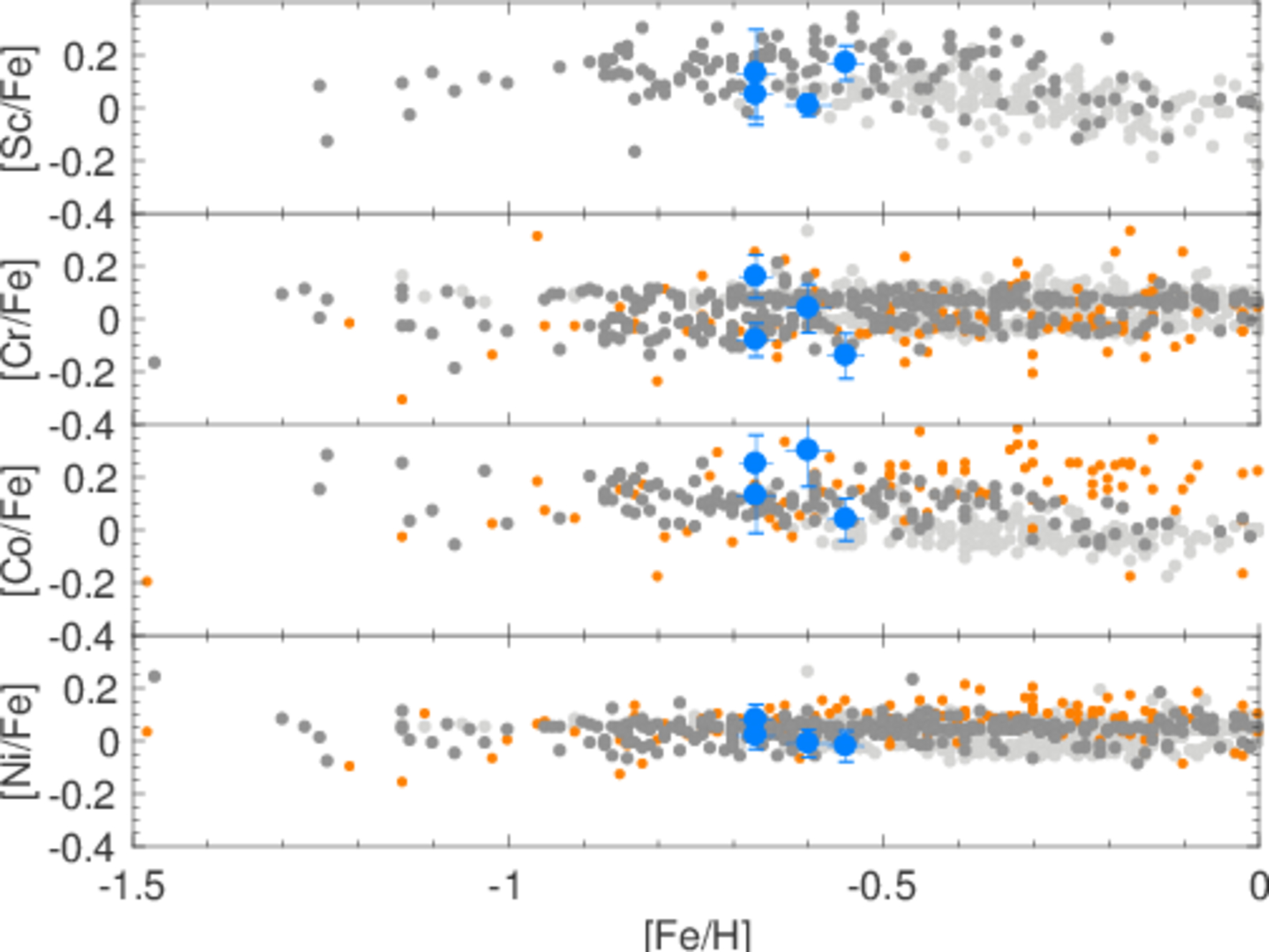}
\caption{Same as Fig.~4, but for the Fe-peak elements.}
\end{figure}

The majority of Fe-peak elements are coproduced in SNe Ia explosions alongside iron and their abundances should reflect this trend, irrespective of their environment. 
As seen in Fig.~5, this is indeed the fact in our Gaia~1 stars, although no sharp division is seen with respect to the Galactic components  for the case of Ni and Cr.
While Fig.~5 seems to imply the presence of a significant spread in the Cr-abundances, there is no obvious physical cause for such a dispersion, also in comparison with 
elements that undergo similar nucleosynthetic formation mechanisms; this rather hints that we underestimated the measurement errors for Cr or small, residual NLTE corrections 
that differ amongst our stars. 

\citet{Nissen2000}  ascertained that Sc behaves like an $\alpha$-element in Galactic disk and halo stars, following a decreasing trend towards higher metallicities in the disks and 
showing a dichotomy in the halo, as also seen in terms of the dual high- and low-$\alpha$ components of the Milky Way \citep{NissenSchuster2010}. 
This general trend is corroborated by our data and an overlap of Gaia~1 with the disk samples is clearly visible, although no clear cut between the various subcomponents beyond the measurement uncertainties
can be established. 

Overall, we are left with the conclusion 
that the material that Gaia~1 formed from was salted by SNe I ejecta to the same amount as typical of its environment, irrespective of its position in the Galaxy, or its age.
\subsection{Neutron-capture elements: Ba, Eu}
In order to derive abundances of the heavy elements  Ba and Eu we resorted to spectral synthesis.
For Ba, we adopted the  error-weighted mean of the  three strong, yet unsaturated, lines at 5853, 6164, and 6496 ~\AA~as 
the final [Ba/Fe] ratio, while the value for  Eu was obtained from the overall weaker feature at 6645 ~\AA.
The results are presented in Fig.~6.
\begin{figure}[htb]
\centering
\includegraphics[width=1\hsize]{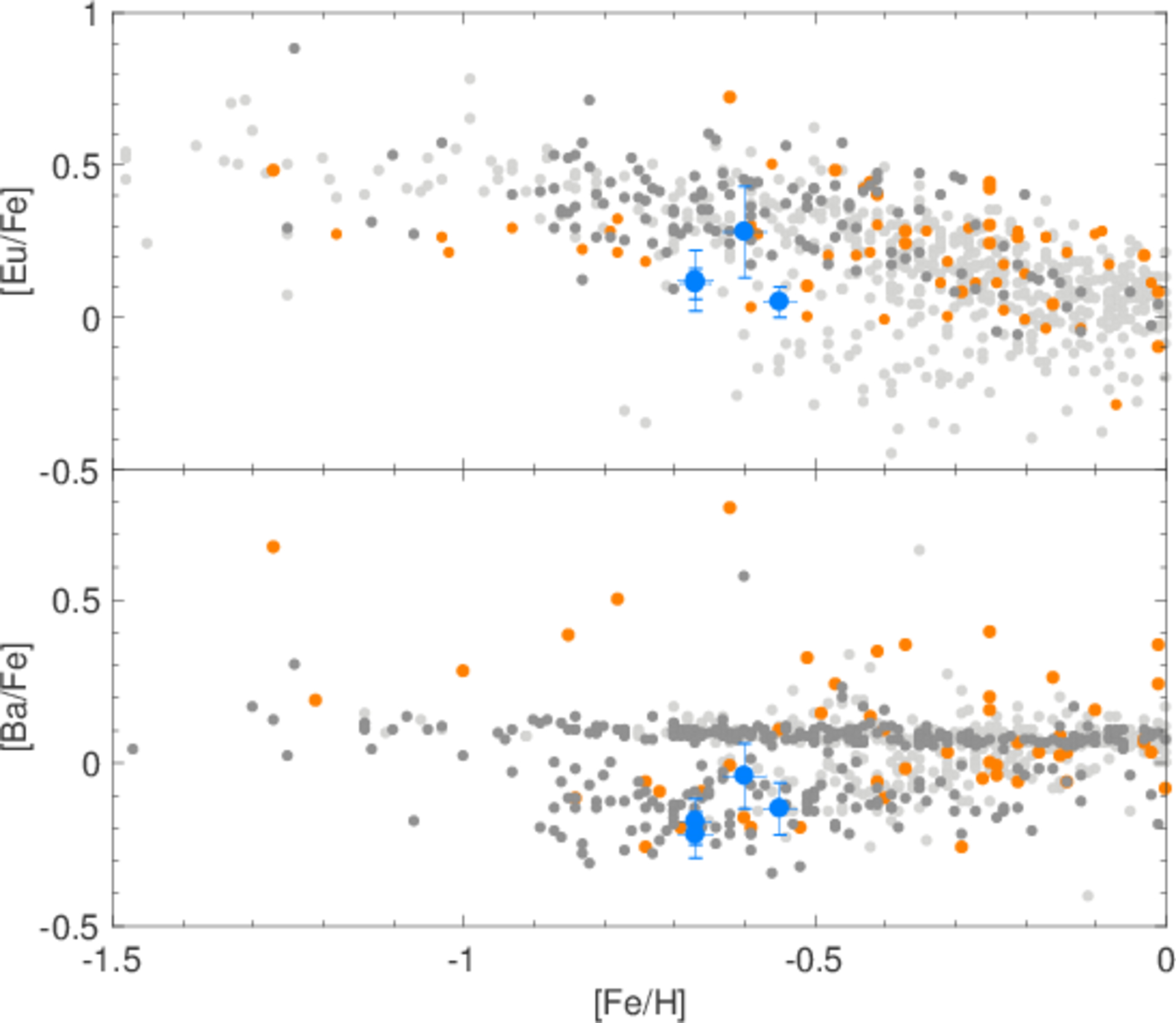}
\caption{Same as Fig.~4 but for the $n$-capture elements Ba and Eu.}
\end{figure}

As an $r$-process element, Eu is primarily produced in type II supernovae (SNe II)  or via neutron star mergers \citep{WoosleyWeaver1995,Freiburghaus1999}.
Given the different timescales involved with those events compared to the slow evolution of SNe Ia that are responsible for the bulk of Fe-production, also 
the [Eu/Fe] ratio sees a decline from elevated values at the metal-poor end to (sub-)solar ratios for the more metal-rich components. 
Gaia~1 has Eu-abundances that follow this trend, albeit skimming the lower edge of the distribution. 

Compared to the $\alpha$-elements the respective Galactic chemical evolution of neutron capture elements is more complex, leading to a larger scatter in the heavy element abundances 
at a given metallicity \citep[e.g.,][]{Sneden2008,CJHansen2014}. Nonetheless, combining the two chemical tracers into a simplistic [($\alpha$+Eu)/Fe] ratio has been shown 
to have a large discriminatory power between the kinematically selected thin and thick disks \citep{Navarro2011}.
This is indicated in Fig.~7, where we overplot our measurements of Gaia~1 on the literature samples as before. 
With the possible exception of star \#2, our targets in the Gaia~1 cluster are highly indicative of a thick disk association, as 
already evidenced by the other chemical tracers discussed above. 
\begin{figure}[htb]
\centering
\includegraphics[width=1\hsize]{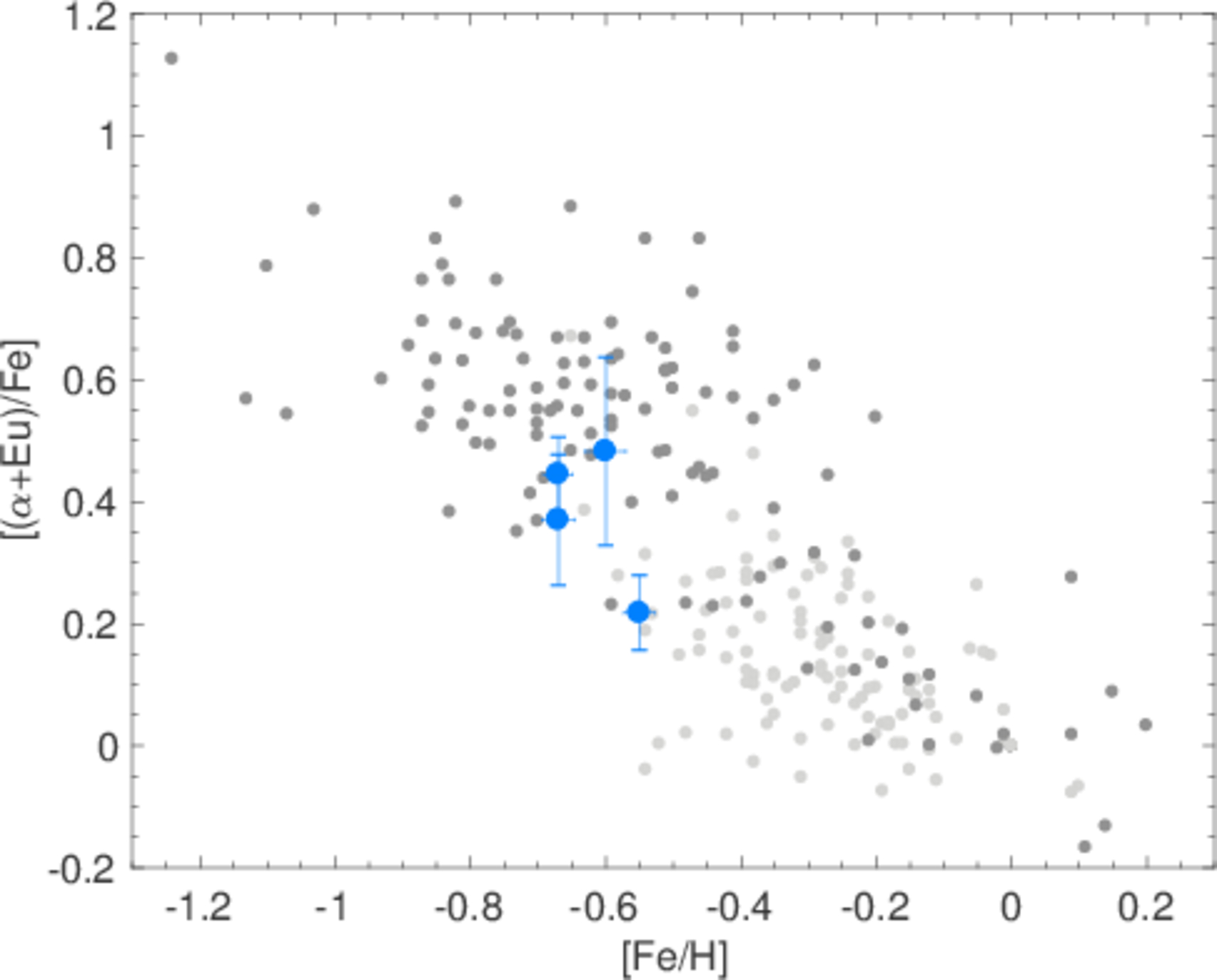}
\caption{Combination of the averaged $\alpha$-element abundances with the $r$-process element Eu, following \citet{Navarro2011}, for the thick and thin disk samples of 
\citet{Reddy2003,Reddy2006}.}
\end{figure}

Barium is produced in the $r$-process at low metallicities, with ever-growing contributions from the $s$-process with increasing metallicities above [Fe/H]$\ga -2.4$
\citep{Busso1999,Simmerer2004,Karakas2014}. 
As with Eu above, the Gaia~1 stars populate the lower branch seen in the [Ba/Fe] abundance distribution (lower panel of Fig.~6). 

\begin{figure}[htb]
\centering
\includegraphics[width=1\hsize]{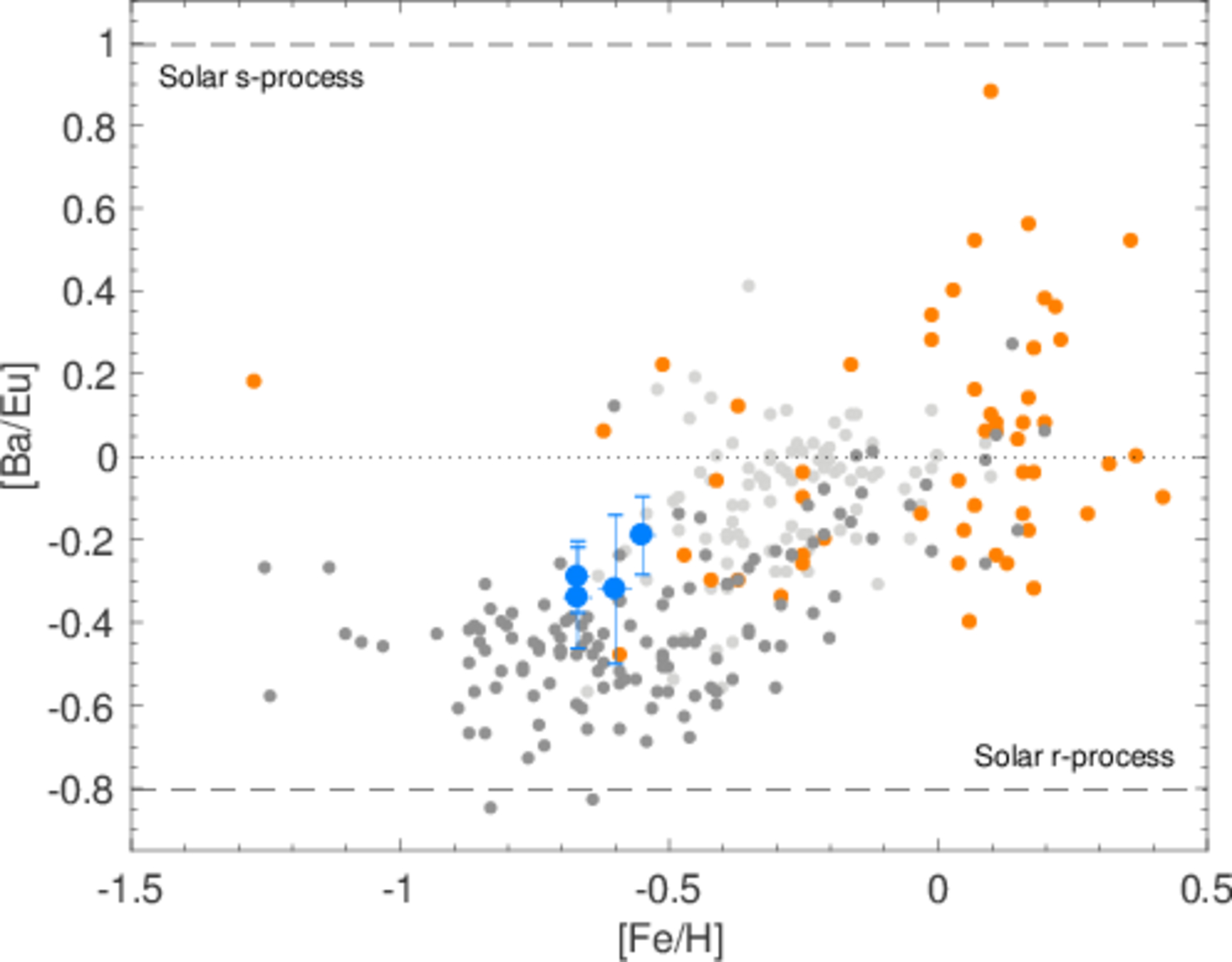}
\caption{The $s$-to-$r$-ratio [Ba/Eu] for the same Galactic samples as shown in the previous figures. 
The Solar values are from  \citet{Simmerer2004}.}
\end{figure}
As is typical for  moderately metal-poor disk stars,  Gaia~1 has experienced some level of $s$-process activity, as is reflected 
in a [Ba/Eu] ratio of $\sim$0.3 in out stars (Fig.~8). This is
consistent with the broad range of this [$s$/$r$] process indicator seen in disk field stars and star clusters at similar metallicities. 
Not surprisingly, Gaia~1 also overlaps with the thick disk ratios in this parameter space, while thin disk and bulge stars, at higher metallicities, 
show progressively higher $s$-process contributions towards the solar value. 
Here we note the interesting exception of the metal-rich ([Fe/H]=$-0.50$ dex) bulge GC NGC~6440, which has recently been exposed as a 
 as a pure $r$-process system at [Ba/Eu]$\sim$0.65 \citep{Munoz2017} with heavy element production being dominated by explosive nucleosynthesis, e.g., in SNe II.
This finding added to the complexity of this Galactic component, leading \citet{Munoz2017} to suggest that the 
chemical enrichment of bulge GCs was special and may have been different from the remainder of the Galactic population. 
\section{Kinematics and orbit -- Gaia 1 in the era of Gaia}
\subsection{Radial velocities and dynamic mass}
Radial velocities were measured from a cross correlation of the heliocentrically corrected spectra against a typical red giant star template, yielding a precision of typically 0.3 km\,s$^{-1}$.
{ The accuracy of the zeropoints in our wavelength scale was ascertained by cross correlating  the telluric A- and B-bands against synthetic atmospheric spectra tailored to the sky conditions 
at the point of our observations, using the TAPAS (Transmissions Atmosph\'eriques Personnalis\'ees Pour l'AStronomie) simulator \citep{Bertaux2014}. 
}

As a result, we established the mean heliocentric radial velocity of Gaia~1 as  57.2$\pm$1.5 km\,s$^{-1}$ with a dispersion of  $\sigma$=3.0$\pm$1.1 km\,s$^{-1}$. 
While the mean value is in excellent agreement with that found 
by both \citet{Simpson2017} and \citet{Mucciarelli2017}, our dispersion is larger by $\sim$2 km\,$^{-1}$.
{ 
Here, it is worth mentioning that, as discussed further in Sect.~5.2, the proper motions of all our four stars clearly associate them 
as cluster members, so there is no indication of a field star inflating the dispersion in velocities.
In fact, this higher velocity dispersion appears to be driven by the higher radial velocity of member star \#3. 
We cannot exclude the possibility that this object is in a binary, but a better temporal coverage of observations would be needed to confirm this.}

{ \citet{Koposov2017} estimated the system's luminous mass (down to $G<19$ mag) to be 2.2$\times$10$^4$ M$_{\odot}$ 
from their radial number density profile, adopting their best-fit isochrone and a 
initial mass function from \citet{Chabrier2003} and assuming that mass follows light.
Using our kinematic data we can attempt an independent measure of the cluster's dynamic mass. 
A tentative value can be gleaned from the formalism of \citet{Illingworth1976} as M$_{\rm dyn}\sim167\mu\sigma^2r_c$, where
$r_c$ refers to the core-radius and $\mu$ is the dimensionless mass, scaling with the object's concentration \citep{King1966}.
By assuming Gaia~1's half-light radius of 9$\pm$0.6 pc as an upper limit for $r_c$ and its full extent from the profile of \citet{Koposov2017}, 
this yields a consistent value of M$_{\rm dyn}=(3^{+4}_{-2})\times10^4$ M$_{\odot}$. 
In turn, using the prescription of
\citet{Spitzer1987} in terms of the better-defined half-light radius, i.e., M$_{\rm dyn}=9.75 r_h \sigma^2 / G$, where $G$ is the gravitational constant, 
yields a larger value of (1.8$\pm$1.4)$\times$10$^5$ M$_{\odot}$. While still broadly consistent with the luminous mass within the uncertainties, 
this value would rather imply the presence of a dark matter component in Gaia~1, more in line with its being a 
dwarf galaxy than a ``simple'' star cluster \citep{Mateo1993,Walker2007,Gilmore2007}. 
Alternatively, Gaia~1 could have been a formerly more massive system that had lost a great deal of its initial mass. This would 
render the occurrence of a spread in Fe and/or multiple populations in this object a viable possibility \citep{Carretta2009Fe,Koch2012,Kirby2013}. 
However, this interpretation hinges on the measured, large velocity dispersion:  adopting a lower value of (0.94$\pm$0.15) km\,s$^{-1}$ \citep{Simpson2017} renders the 
mass lower by a factor of ten, at (1.8$\pm$0.6)$\times$10$^4$ M$_{\odot}$. Given the low number of targets involved, we will not place further weight on 
these mass arguments. 

Overall, within the error our finding is compatible 
with the low dispersions found in other GCs of comparable mass. Here, we note that the works of \citet{Pryor1993} and \citet[][2010 edition]{Harris1996} only list one object, Pal~13, that is fainter than Gaia~1 with a measured velocity dispersion \citep{Cote2002,Blecha2004}.
Similarly, OCs exhibit low intrinsic dispersions \citep{Friel2013} below 10 km\,s$^{-1}$, with a tendency of increasing values for older systems \citep{Hayes2014}. 

}

\subsection{Proper motions and orbit}
As in our previous work on largely uncharted star clusters \citep{Koch2017ESO}, we computed the 
orbits of our four target stars using proper motions from the UCAC5 catalog \citep{Zacharias2017}. 
Using the {\it Gaia}-TGAS stars in the 8 to 11
magnitude range a reference star catalog, the UCAC5 catalog released new proper motions 
for the high quality, ground-based US Naval Observatory CCD Astrograph Catalog (UCAC) 
all-sky observations on the Gaia coordinate system.  This allows the UCAC5 proper 
motions to be similar in performance to the {\it Gaia}-TGAS ones, but for a much larger
sample of stars -- over 107 million stars have proper 
motion measurements with typical accuracies of 1 to 2 $\rm mas~yr^{-1}$.
The UCAC5 proper motions are, in general, improved compared those presented in TGAS. 
Fig.~9 shows the UCAC5 proper motions for all UCAC5 sources centred on Gaia~1, 
within a 7.5 arcmin radius.  The stars presented here, as well as those observed by \citet{Simpson2017} and 
\citet{Mucciarelli2017}, are also shown.  It is clear that the proper motion estimated by 
\citet{Simpson2017} -- by combining the {\it Gaia} and 2MASS positional information -- is offset 
from the mean UCAC5 proper 
motions of the cluster stars.  All of our stars have proper motions consistent with cluster membership, as
do the \citet{Mucciarelli2017} stars, albeit their star \#3 which they discarded due to damped absorption lines
and a marginally different radial velocity.
\begin{figure}[htb]
\centering
\includegraphics[width=1\hsize]{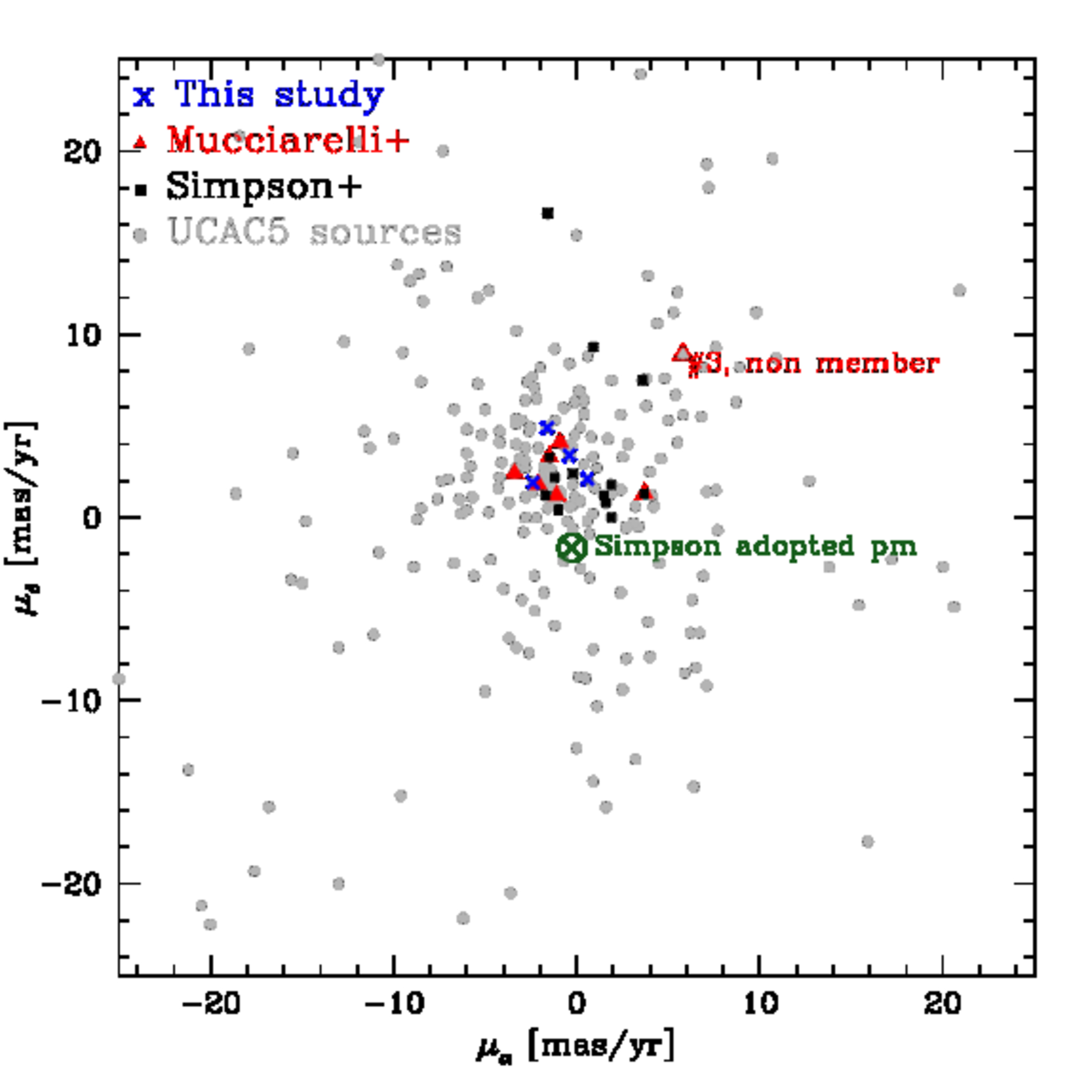}
\caption{Proper motions of stars within 7.5$\arcmin$ of Gaia~1. Spectroscopic targets from \citet{Simpson2017}, \citet{Mucciarelli2017}, and our work are highlighted.}
\end{figure}

Owing to our spectroscopically derived gravities, we derived distances to our stars, further adopting isochrones 
that reflect our measured metallicites. 
The resulting heliocentric distance of 4.1$\pm$1.1 kpc is broadly consistent with the value found 
by \citet{Koposov2017} of 4.6 kpc from sparse photometry.
The derived orbits below  depend only mildly on the distance adopted and the errors in orbital parameters
are chiefly driven by proper motion errors.  
Our star \#4 has the largest formal proper motion uncertainty of 1.9~mas~yr$^{-1}$, whereas the others have
proper motion uncertainties between 1.2 and 1.4~mas~yr$^{-1}$.  

Using the {\tt galpy} Python package\footnote{http://github.com/jobovy/galpy; version 1.2} 
and the recommended {\tt MWPotential2014} potential
with the default parameters \citep{Bovy2015}, 
we integrated the orbit backwards in time for 10 Gyr.  The resulting orbit projections are shown in Fig.~9. 
As expected for open clusters, the stars show quasi-periodic crown orbits.
\begin{figure}[htb]
\centering
\includegraphics[width=1\hsize]{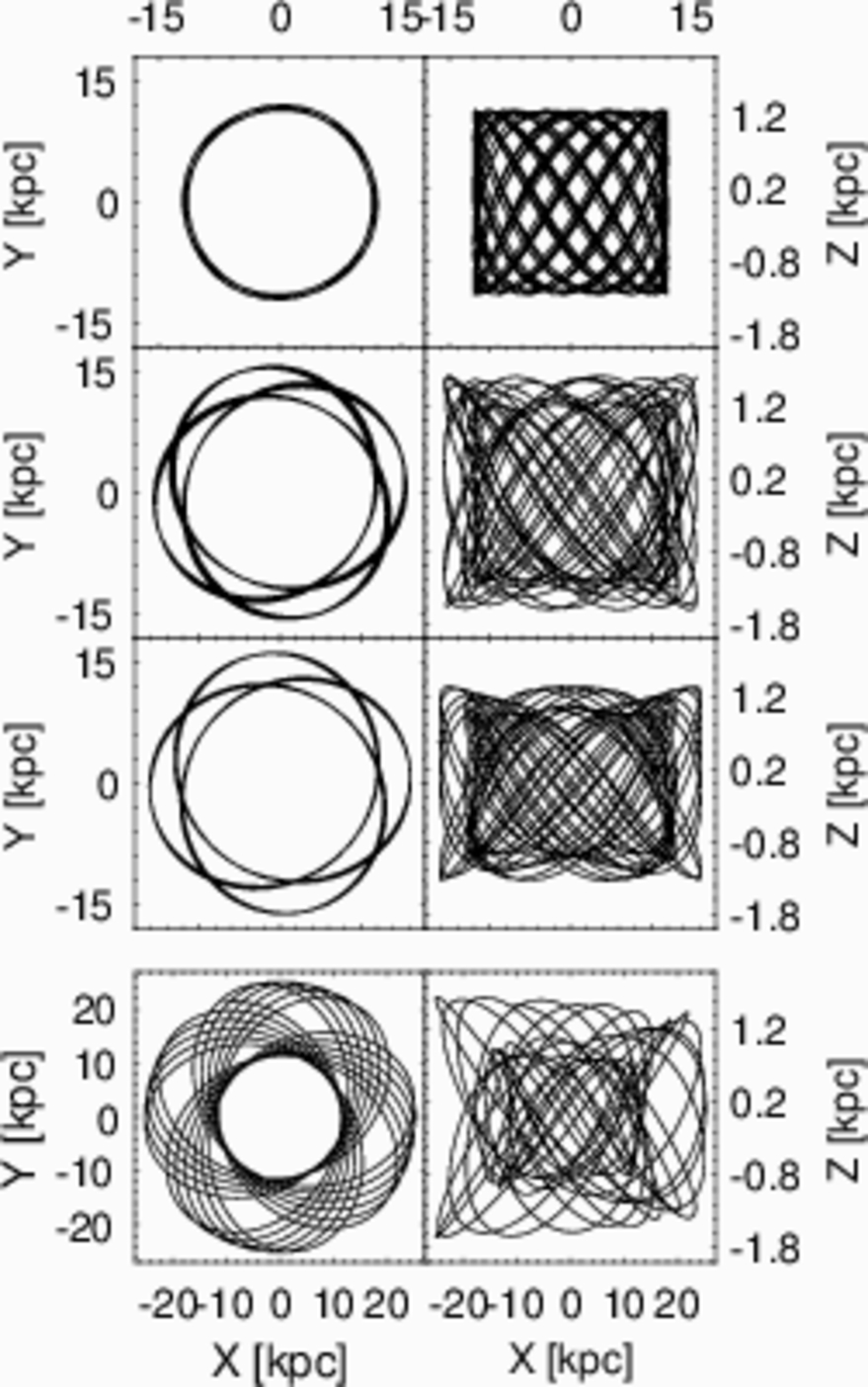}
\caption{Orbit projection of stars \#1--4 (top to bottom).}
\end{figure}

The mean eccentricity of our stars is 0.14, which is an ellipticity value that coincides with the peak
of thick-disk eccentricity distribution from $\sim$30 000 thick disk stars presented in \citet{Dierickx2010}. 
It is, however, also consistent with a thin disk eccentricity, as the range between 0.1 to 0.3 
is the eccentricity regime in which the thin and thick-disk stars overlap \citep{Lee2011,Boeche2013}. 
Our eccentricity gives Gaia 1 a considerably rounder orbit
then that derived by \citet{Simpson2017}.  The eccentricity is largely insensitive to the actual 
distance within a 0.5~kpc range, so the factor of two difference between the eccentricity
found here and that presented in \citet{Simpson2017} is almost entirely due to the different
proper motions used (see also Figure~9).

Importantly, because our orbital ellipticity is half the size of the ellipticity calculated by \citet{Simpson2017},
the radial quantity $\eta$ \citep[see also][]{VandePutte2010} invoked by \citet{Simpson2017} to argue for 
an extra-galactic origin of this cluster, is also half the size as that found by these authors.  We find
$\eta$ = 0.24, which is well inline with a typical open cluster, especially considering
it's metallicity, as more metal-poor open clusters tend to have larger $\eta$ values.
Moreover, the maximum height above the Galactic plane, $Z_{\rm max}$ resulting from our integration 
reaches $\sim$1.0 kpc for the four stars, 
which is in line with the scale height of the thick disk, but in contrast to the larger
value of 1.7 kpc found by those authors.  Unfortunately the $Z_{\rm max}$ depends more 
strongly on the adopted distance as compared to eccentricity, so this parameter is more uncertain. 
Still, taken together with the above discussion, we do not find strong evidence that Gaia 1 has 
a particularly anomalous orbit.

Fig.~11 shows a Toomre diagram of Galactic stars (left panel) that have been kinematically tagged as thin disk, thick disk, or halo stars by \citet{Reddy2003,Reddy2006} and \citet{Bensby2014}. 
Similarly, the right panel includes kinematic measurements of Galactic OCs \citep{VandePutte2010} that were divided into the Galactic components following the prescription of 
\citet{Bensby2014}. 
\begin{figure*}[htb]
\centering
\includegraphics[width=0.65\hsize]{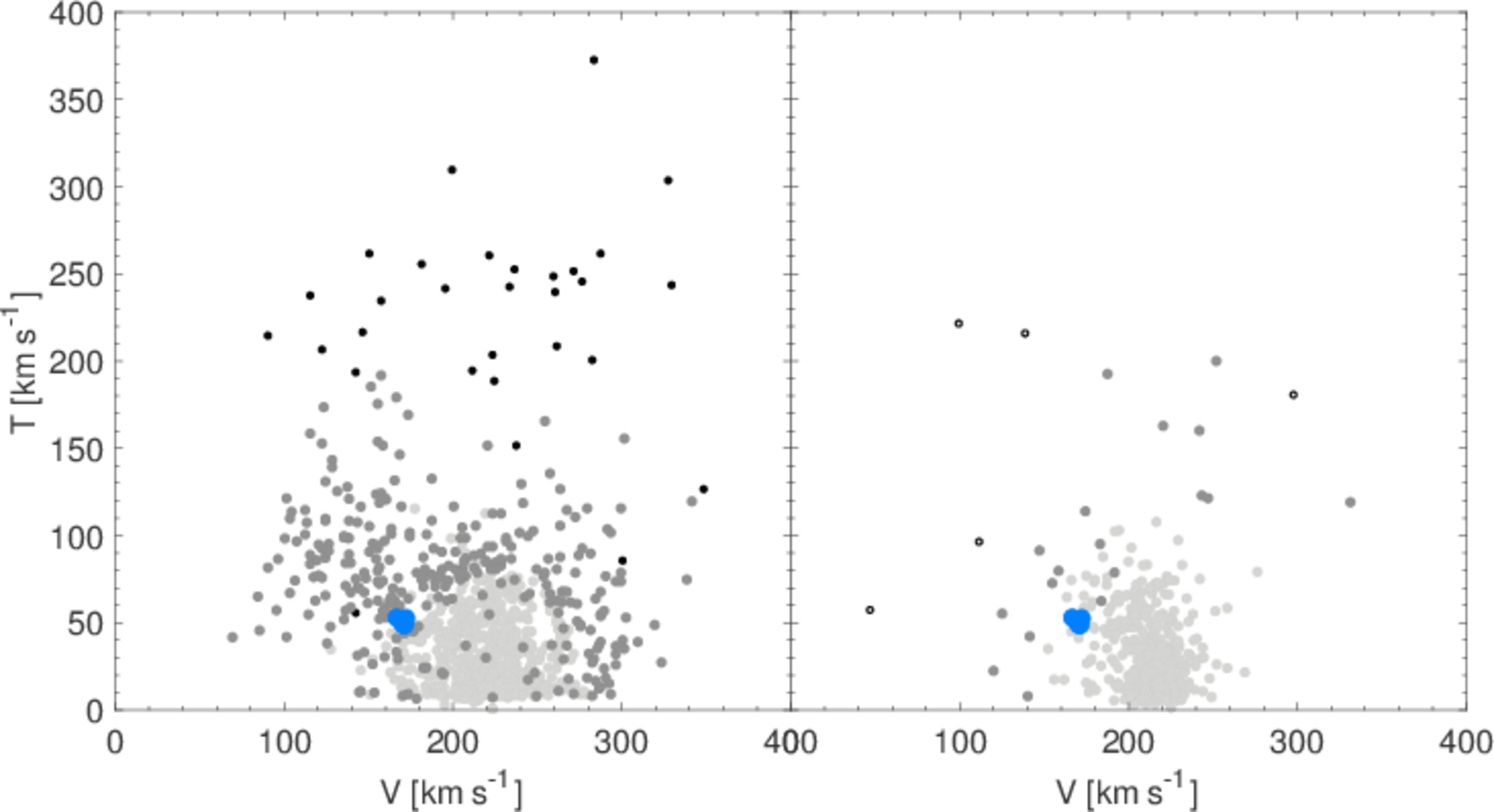}
\caption{Left panel: Toomre diagram of the kinematically separated literature samples and the Gaia~1 stars. 
As in the previous figures, light gray points are thin disk stars, 
dark gray ones are for the thick disk, and black symbols denote stars on halo-like orbits.
The right panel uses OCs from the compilation of \citet{VandePutte2010} as a comparison sample, colour-coded following the decomposition procedure of \citet{Bensby2014}.}
\end{figure*}
While also the OC connection with a thick disk seems feasible, 
\citet{VandePutte2010} assigned properties to their kinematic cluster sample  
based on orbital eccentricity and maximum height above the plane, which are likely correlated with age and 
metallicity of the systems. 
As such, Gaia~1 would qualify as a system that either was accreted into the disk from an extragalactic source, or it resulted from a merger of proto-cluster material with high-latitude gas clouds. 
This has already been suggested by \citep{Simpson2017}, although an extragalactic origin has been excluded by the chemical abundance study of \citet{Mucciarelli2017}.  
{ However, as our present work opened up the possibility of an abundance spread in this system, such an origin cannot fully be discarded anymore.}
Given that only $\sim$5\% of open clusters
are associated with the thick disk \citep{Holmberg2007,Wu2009}, the statistical samples to understand these systems
are just not as large.
 Our chemo-dynamical tagging of Gaia~1 was finally able to ascertain its present association with the thick disk, whereas its exact formation and origin remains ambiguous.  
\section{Conclusions}
Chemical tagging of stellar populations has proven indispensable  when the underlying details of the cluster in question are sparse. Even the addition of a handful of spectroscopic measurements of candidate member stars can often establish  a unique association of an object with either of the Galactic components. This was recently exemplified by the low-mass GC 
ESO452-Sc11 \citep{Bica1999},  which \citet{Koch2017ESO} chemically tied to the Galactic bulge. 

Likewise, settling an accurate abundance scale for stellar systems is imperative for deriving their ages. 
In this context, views about the nature of the target of the present work, Gaia~1, are ambivalent.  This is also in part because the distance to the cluster is poorly constrained -- 
for example, the main sequence turn-off, which has considerable power to discriminate between isochrones, has not clearly been imaged yet.
The discovery of Gaia~1 implied an intermediate isochrone-age and metallicity of 6.3 Gyr and $-0.7$ dex \citep{Koposov2017}. In contrast, first spectroscopic analyses \citep{Simpson2017,Mucciarelli2017} 
led to an  age of 3 Gyr based on  CaT  and  EW measurements plus spectral synthesis at similar resolution to ours, albeit with smaller wavelength coverage.
The resulting [Fe/H] was higher by more than half a dex than our values from a high-resolution abundance analysis, which returned metallicities in line with the { preliminary} colour-magnitude diagram fitting. 

Our abundance analysis revealed $\alpha$-abundances that are are consistent with the thick disk, consolidated by the neutron-capture elements and finally bolstered by 
the stars' kinematics. Despite the presence of 
mild light element variations in Na, Mg, and Al no obvious (anti-)correlations are seen between those elements as would be expected in Galactic GCs. 
Since its stellar mass of a few ten thousand M$_{\odot}$ places 
Gaia~1 at the boundary between very low-mass GCs and very luminous OCs, our analysis rather comforts the latter view,  
 as already suggested by the CMD study of \citet{Koposov2017}.
 Thus, our chemo-dynamical tagging of Gaia~1 enabled us to clearly ascertain its present association with the thick disk
 and rather being a massive OC than a low-mass GC. 
{  However, the hint of a metallicity spread between different studies in the literature may point towards a more complex origin that could involve a once more massive progenitor. 
 Thus} the question as to its exact formation and origin remains unclear and needs to await more data such as 
the precise and accurate parallaxes that {\em Gaia} can offer. 
\begin{acknowledgements}
We are grateful to the referee, Piercarlo Bonifacio, for a very helpful and thorough report that 
helped to better understand the discrepancies with the literature. 
This work was supported in parts by Sonderforschungsbereich SFB 881 "The Milky Way System" (subproject A8) 
of the German Research Foundation (DFG).
This work has made use of data from the European Space Agency (ESA)
mission {\it Gaia} ({https://www.cosmos.esa.int/gaia}), processed by
the {\it Gaia} Data Processing and Analysis Consortium (DPAC,
l{https://www.cosmos.esa.int/web/gaia/dpac/consortium}). Funding
for the DPAC has been provided by national institutions, in particular
the institutions participating in the {\it Gaia} Multilateral Agreement.
\end{acknowledgements}
\bibliographystyle{aa} % style aa.bst
\bibliography{ms} % your references Yourfile.bib

\begin{thebibliography}{100}
\expandafter\ifx\csname natexlab\endcsname\relax\def\natexlab#1{#1}\fi

\bibitem[{{Alonso} {et~al.}(1999){Alonso}, {Arribas}, \&
  {Mart{\'{\i}}nez-Roger}}]{Alonso1999}
{Alonso}, A., {Arribas}, S., \& {Mart{\'{\i}}nez-Roger}, C. 1999, \aaps, 140,
  261

\bibitem[{{Asplund} {et~al.}(2009){Asplund}, {Grevesse}, {Sauval}, \&
  {Scott}}]{Asplund2009}
{Asplund}, M., {Grevesse}, N., {Sauval}, A.~J., \& {Scott}, P. 2009, \araa, 47,
  481

\bibitem[{{Battistini} \& {Bensby}(2016)}]{Battistini2016}
{Battistini}, C. \& {Bensby}, T. 2016, \aap, 586, A49

\bibitem[{{Bensby} {et~al.}(2014){Bensby}, {Feltzing}, \& {Oey}}]{Bensby2014}
{Bensby}, T., {Feltzing}, S., \& {Oey}, M.~S. 2014, \aap, 562, A71

\bibitem[{{Bensby} {et~al.}(2013){Bensby}, {Yee}, {Feltzing}, {Johnson},
  {Gould}, {Cohen}, {Asplund}, {Mel{\'e}ndez}, {Lucatello}, {Han}, {Thompson},
  {Gal-Yam}, {Udalski}, {Bennett}, {Bond}, {Kohei}, {Sumi}, {Suzuki}, {Suzuki},
  {Takino}, {Tristram}, {Yamai}, \& {Yonehara}}]{Bensby2013}
{Bensby}, T., {Yee}, J.~C., {Feltzing}, S., {et~al.} 2013, \aap, 549, A147

\bibitem[{{Bergemann} \& {Cescutti}(2010)}]{Bergemann2010Cr}
{Bergemann}, M. \& {Cescutti}, G. 2010, \aap, 522, A9

\bibitem[{{Bergemann} {et~al.}(2010){Bergemann}, {Pickering}, \&
  {Gehren}}]{Bergemann2010Co}
{Bergemann}, M., {Pickering}, J.~C., \& {Gehren}, T. 2010, \mnras, 401, 1334

\bibitem[{{Bertaux} {et~al.}(2014){Bertaux}, {Lallement}, {Ferron}, {Boonne},
  \& {Bodichon}}]{Bertaux2014}
{Bertaux}, J.~L., {Lallement}, R., {Ferron}, S., {Boonne}, C., \& {Bodichon},
  R. 2014, \aap, 564, A46

\bibitem[{{Bica} {et~al.}(1999){Bica}, {Ortolani}, \& {Barbuy}}]{Bica1999}
{Bica}, E., {Ortolani}, S., \& {Barbuy}, B. 1999, \aaps, 136, 363

\bibitem[{{Blecha} {et~al.}(2004){Blecha}, {Meylan}, {North}, \&
  {Royer}}]{Blecha2004}
{Blecha}, A., {Meylan}, G., {North}, P., \& {Royer}, F. 2004, \aap, 419, 533

\bibitem[{{Boeche} {et~al.}(2013){Boeche}, {Chiappini}, {Minchev}, {Williams},
  {Steinmetz}, {Sharma}, {Kordopatis}, {Bland-Hawthorn}, {Bienaym{\'e}},
  {Gibson}, {Gilmore}, {Grebel}, {Helmi}, {Munari}, {Navarro}, {Parker},
  {Reid}, {Seabroke}, {Siebert}, {Siviero}, {Watson}, {Wyse}, \&
  {Zwitter}}]{Boeche2013}
{Boeche}, C., {Chiappini}, C., {Minchev}, I., {et~al.} 2013, \aap, 553, A19

\bibitem[{{Bovy}(2015)}]{Bovy2015}
{Bovy}, J. 2015, \apjs, 216, 29

\bibitem[{{Bragaglia} {et~al.}(2012){Bragaglia}, {Gratton}, {Carretta},
  {D'Orazi}, {Sneden}, \& {Lucatello}}]{Bragaglia2012}
{Bragaglia}, A., {Gratton}, R.~G., {Carretta}, E., {et~al.} 2012, \aap, 548,
  A122

\bibitem[{{Busso} {et~al.}(1999){Busso}, {Gallino}, \&
  {Wasserburg}}]{Busso1999}
{Busso}, M., {Gallino}, R., \& {Wasserburg}, G.~J. 1999, \araa, 37, 239

\bibitem[{{Carretta} {et~al.}(2009{\natexlab{a}}){Carretta}, {Bragaglia},
  {Gratton}, {D'Orazi}, \& {Lucatello}}]{Carretta2009Fe}
{Carretta}, E., {Bragaglia}, A., {Gratton}, R., {D'Orazi}, V., \& {Lucatello},
  S. 2009{\natexlab{a}}, \aap, 508, 695

\bibitem[{{Carretta} {et~al.}(2009{\natexlab{b}}){Carretta}, {Bragaglia},
  {Gratton}, \& {Lucatello}}]{Carretta2009NaO}
{Carretta}, E., {Bragaglia}, A., {Gratton}, R., \& {Lucatello}, S.
  2009{\natexlab{b}}, \aap, 505, 139

\bibitem[{{Carretta} {et~al.}(2010){Carretta}, {Bragaglia}, {Gratton},
  {Recio-Blanco}, {Lucatello}, {D'Orazi}, \& {Cassisi}}]{Carretta2010}
{Carretta}, E., {Bragaglia}, A., {Gratton}, R.~G., {et~al.} 2010, \aap, 516,
  A55

\bibitem[{{Castelli} \& {Kurucz}(2004)}]{CastelliKurucz2004}
{Castelli}, F. \& {Kurucz}, R.~L. 2004, ArXiv Astrophysics e-prints

\bibitem[{{Chabrier}(2003)}]{Chabrier2003}
{Chabrier}, G. 2003, \pasp, 115, 763

\bibitem[{{Chambers} {et~al.}(2016){Chambers}, {Magnier}, {Metcalfe},
  {Flewelling}, {Huber}, {Waters}, {Denneau}, {Draper}, {Farrow}, {Finkbeiner},
  {Holmberg}, {Koppenhoefer}, {Price}, {Saglia}, {Schlafly}, {Smartt},
  {Sweeney}, {Wainscoat}, {Burgett}, {Grav}, {Heasley}, {Hodapp}, {Jedicke},
  {Kaiser}, {Kudritzki}, {Luppino}, {Lupton}, {Monet}, {Morgan}, {Onaka},
  {Stubbs}, {Tonry}, {Banados}, {Bell}, {Bender}, {Bernard}, {Botticella},
  {Casertano}, {Chastel}, {Chen}, {Chen}, {Cole}, {Deacon}, {Frenk},
  {Fitzsimmons}, {Gezari}, {Goessl}, {Goggia}, {Goldman}, {Grebel}, {Hambly},
  {Hasinger}, {Heavens}, {Heckman}, {Henderson}, {Henning}, {Holman}, {Hopp},
  {Ip}, {Isani}, {Keyes}, {Koekemoer}, {Kotak}, {Long}, {Lucey}, {Liu},
  {Martin}, {McLean}, {Morganson}, {Murphy}, {Nieto-Santisteban}, {Norberg},
  {Peacock}, {Pier}, {Postman}, {Primak}, {Rae}, {Rest}, {Riess}, {Riffeser},
  {Rix}, {Roser}, {Schilbach}, {Schultz}, {Scolnic}, {Szalay}, {Seitz},
  {Shiao}, {Small}, {Smith}, {Soderblom}, {Taylor}, {Thakar}, {Thiel},
  {Thilker}, {Urata}, {Valenti}, {Walter}, {Watters}, {Werner}, {White},
  {Wood-Vasey}, \& {Wyse}}]{Chambers2016}
{Chambers}, K.~C., {Magnier}, E.~A., {Metcalfe}, N., {et~al.} 2016, ArXiv
  e-prints

\bibitem[{{Chen} {et~al.}(2003){Chen}, {Hou}, \& {Wang}}]{Chen2003}
{Chen}, L., {Hou}, J.~L., \& {Wang}, J.~J. 2003, \aj, 125, 1397

\bibitem[{{C{\^o}t{\'e}} {et~al.}(2002){C{\^o}t{\'e}}, {Djorgovski}, {Meylan},
  {Castro}, \& {McCarthy}}]{Cote2002}
{C{\^o}t{\'e}}, P., {Djorgovski}, S.~G., {Meylan}, G., {Castro}, S., \&
  {McCarthy}, J.~K. 2002, \apj, 574, 783

\bibitem[{{Cunha} {et~al.}(2016){Cunha}, {Frinchaboy}, {Souto}, {Thompson},
  {Zasowski}, {Allende Prieto}, {Carrera}, {Chiappini}, {Donor},
  {Garc{\'{\i}}a-Hern{\'a}ndez}, {Garc{\'{\i}}a P{\'e}rez}, {Hayden},
  {Holtzman}, {Jackson}, {Johnson}, {Majewski}, {M{\'e}sz{\'a}ros}, {Meyer},
  {Nidever}, {O'Connell}, {Schiavon}, {Schultheis}, {Shetrone}, {Simmons},
  {Smith}, \& {et al.}}]{Cunha2016}
{Cunha}, K., {Frinchaboy}, P.~M., {Souto}, D., {et~al.} 2016, Astronomische
  Nachrichten, 337, 922

\bibitem[{{Cutri} {et~al.}(2003){Cutri}, {Skrutskie}, {van Dyk}, {Beichman},
  {Carpenter}, {Chester}, {Cambresy}, {Evans}, {Fowler}, {Gizis}, {Howard},
  {Huchra}, {Jarrett}, {Kopan}, {Kirkpatrick}, {Light}, {Marsh}, {McCallon},
  {Schneider}, {Stiening}, {Sykes}, {Weinberg}, {Wheaton}, {Wheelock}, \&
  {Zacarias}}]{Cutri2003}
{Cutri}, R.~M., {Skrutskie}, M.~F., {van Dyk}, S., {et~al.} 2003, {2MASS All
  Sky Catalog of point sources.}

\bibitem[{{Dierickx} {et~al.}(2010){Dierickx}, {Klement}, {Rix}, \&
  {Liu}}]{Dierickx2010}
{Dierickx}, M., {Klement}, R., {Rix}, H.-W., \& {Liu}, C. 2010, \apjl, 725,
  L186

\bibitem[{{Dotter} {et~al.}(2008){Dotter}, {Chaboyer}, {Jevremovi{\'c}},
  {Kostov}, {Baron}, \& {Ferguson}}]{Dotter2008}
{Dotter}, A., {Chaboyer}, B., {Jevremovi{\'c}}, D., {et~al.} 2008, \apjs, 178,
  89

\bibitem[{{Feltzing} \& {Chiba}(2013)}]{Feltzing2013}
{Feltzing}, S. \& {Chiba}, M. 2013, \nar, 57, 80

\bibitem[{{Freiburghaus} {et~al.}(1999){Freiburghaus}, {Rosswog}, \&
  {Thielemann}}]{Freiburghaus1999}
{Freiburghaus}, C., {Rosswog}, S., \& {Thielemann}, F.-K. 1999, \apjl, 525,
  L121

\bibitem[{{Friel}(2013)}]{Friel2013}
{Friel}, E.~D. 2013, {Open Clusters and Their Role in the Galaxy}, ed. T.~D.
  {Oswalt} \& G.~{Gilmore}, 347

\bibitem[{{Geisler} {et~al.}(2007){Geisler}, {Wallerstein}, {Smith}, \&
  {Casetti-Dinescu}}]{Geisler2007}
{Geisler}, D., {Wallerstein}, G., {Smith}, V.~V., \& {Casetti-Dinescu}, D.~I.
  2007, \pasp, 119, 939

\bibitem[{{Gilmore} {et~al.}(2007){Gilmore}, {Wilkinson}, {Wyse}, {Kleyna},
  {Koch}, {Evans}, \& {Grebel}}]{Gilmore2007}
{Gilmore}, G., {Wilkinson}, M.~I., {Wyse}, R.~F.~G., {et~al.} 2007, \apj, 663,
  948

\bibitem[{{Gonzalez} {et~al.}(2011){Gonzalez}, {Rejkuba}, {Zoccali}, {Hill},
  {Battaglia}, {Babusiaux}, {Minniti}, {Barbuy}, {Alves-Brito}, {Renzini},
  {Gomez}, \& {Ortolani}}]{Gonzalez2011}
{Gonzalez}, O.~A., {Rejkuba}, M., {Zoccali}, M., {et~al.} 2011, \aap, 530, A54

\bibitem[{{Hanke} {et~al.}(2017){Hanke}, {Koch}, {Hansen}, \&
  {McWilliam}}]{Hanke2017}
{Hanke}, M., {Koch}, A., {Hansen}, C.~J., \& {McWilliam}, A. 2017, \aap, 599,
  A97

\bibitem[{{Hansen} {et~al.}(2014){Hansen}, {Montes}, \&
  {Arcones}}]{CJHansen2014}
{Hansen}, C.~J., {Montes}, F., \& {Arcones}, A. 2014, \apj, 797, 123

\bibitem[{{Harris}(1996)}]{Harris1996}
{Harris}, W.~E. 1996, \aj, 112, 1487

\bibitem[{{Hayes} \& {Friel}(2014)}]{Hayes2014}
{Hayes}, C.~R. \& {Friel}, E.~D. 2014, \aj, 147, 69

\bibitem[{{Holmberg} {et~al.}(2007){Holmberg}, {Nordstr{\"o}m}, \&
  {Andersen}}]{Holmberg2007}
{Holmberg}, J., {Nordstr{\"o}m}, B., \& {Andersen}, J. 2007, \aap, 475, 519

\bibitem[{{Illingworth}(1976)}]{Illingworth1976}
{Illingworth}, G. 1976, \apj, 204, 73

\bibitem[{{Janes}(1979)}]{Janes1979}
{Janes}, K.~A. 1979, \apjs, 39, 135

\bibitem[{{Johnson} \& {Pilachowski}(2010)}]{Johnson2010}
{Johnson}, C.~I. \& {Pilachowski}, C.~A. 2010, \apj, 722, 1373

\bibitem[{{Johnson} {et~al.}(2012){Johnson}, {Rich}, {Kobayashi}, \&
  {Fulbright}}]{Johnson2012}
{Johnson}, C.~I., {Rich}, R.~M., {Kobayashi}, C., \& {Fulbright}, J.~P. 2012,
  \apj, 749, 175

\bibitem[{{Johnson} {et~al.}(2014){Johnson}, {Rich}, {Kobayashi}, {Kunder}, \&
  {Koch}}]{Johnson2014}
{Johnson}, C.~I., {Rich}, R.~M., {Kobayashi}, C., {Kunder}, A., \& {Koch}, A.
  2014, \aj, 148, 67

\bibitem[{{Karakas} \& {Lattanzio}(2014)}]{Karakas2014}
{Karakas}, A.~I. \& {Lattanzio}, J.~C. 2014, \pasa, 31, e030

\bibitem[{{Kelson}(1998)}]{Kelson1998}
{Kelson}, D.~D. 1998, PhD thesis, , Univ.~California at Santa Cruz, (1998)

\bibitem[{{Kelson}(2003)}]{Kelson2003}
{Kelson}, D.~D. 2003, \pasp, 115, 688

\bibitem[{{Kelson} {et~al.}(2000){Kelson}, {Illingworth}, {van Dokkum}, \&
  {Franx}}]{Kelson2000}
{Kelson}, D.~D., {Illingworth}, G.~D., {van Dokkum}, P.~G., \& {Franx}, M.
  2000, \apj, 531, 159

\bibitem[{{King}(1966)}]{King1966}
{King}, I.~R. 1966, \aj, 71, 64

\bibitem[{{Kirby} {et~al.}(2013){Kirby}, {Cohen}, {Guhathakurta}, {Cheng},
  {Bullock}, \& {Gallazzi}}]{Kirby2013}
{Kirby}, E.~N., {Cohen}, J.~G., {Guhathakurta}, P., {et~al.} 2013, \apj, 779,
  102

\bibitem[{{Koch} \& {Edvardsson}(2002)}]{Koch2002}
{Koch}, A. \& {Edvardsson}, B. 2002, \aap, 381, 500

\bibitem[{{Koch} {et~al.}(2017){Koch}, {Hansen}, \& {Kunder}}]{Koch2017ESO}
{Koch}, A., {Hansen}, C.~J., \& {Kunder}, A. 2017, \aap, 604, A41

\bibitem[{{Koch} {et~al.}(2012){Koch}, {L{\'e}pine}, \& {{\c C}al{\i}{\c
  s}kan}}]{Koch2012}
{Koch}, A., {L{\'e}pine}, S., \& {{\c C}al{\i}{\c s}kan}, {\c S}. 2012, in
  European Physical Journal Web of Conferences, Vol.~19, European Physical
  Journal Web of Conferences, 03002

\bibitem[{{Koch} \& {McWilliam}(2008)}]{Koch2008}
{Koch}, A. \& {McWilliam}, A. 2008, \aj, 135, 1551

\bibitem[{{Koch} \& {McWilliam}(2014)}]{Koch5897}
{Koch}, A. \& {McWilliam}, A. 2014, \aap, 565, A23

\bibitem[{{Koch} {et~al.}(2016){Koch}, {McWilliam}, {Preston}, \&
  {Thompson}}]{Koch2016}
{Koch}, A., {McWilliam}, A., {Preston}, G.~W., \& {Thompson}, I.~B. 2016, \aap,
  587, A124

\bibitem[{{Koposov} {et~al.}(2017){Koposov}, {Belokurov}, \&
  {Torrealba}}]{Koposov2017}
{Koposov}, S.~E., {Belokurov}, V., \& {Torrealba}, G. 2017, \mnras, 470, 2702

\bibitem[{{Lee} {et~al.}(2009){Lee}, {Kang}, {Lee}, \& {Lee}}]{Lee2009}
{Lee}, J.-W., {Kang}, Y.-W., {Lee}, J., \& {Lee}, Y.-W. 2009, \nat, 462, 480

\bibitem[{{Lee} {et~al.}(2011){Lee}, {Beers}, {An}, {Ivezi{\'c}}, {Just},
  {Rockosi}, {Morrison}, {Johnson}, {Sch{\"o}nrich}, {Bird}, {Yanny},
  {Harding}, \& {Rocha-Pinto}}]{Lee2011}
{Lee}, Y.~S., {Beers}, T.~C., {An}, D., {et~al.} 2011, \apj, 738, 187

\bibitem[{{Lind} {et~al.}(2011){Lind}, {Asplund}, {Barklem}, \&
  {Belyaev}}]{Lind2011}
{Lind}, K., {Asplund}, M., {Barklem}, P.~S., \& {Belyaev}, A.~K. 2011, \aap,
  528, A103

\bibitem[{{Lind} {et~al.}(2009){Lind}, {Primas}, {Charbonnel}, {Grundahl}, \&
  {Asplund}}]{Lind2009}
{Lind}, K., {Primas}, F., {Charbonnel}, C., {Grundahl}, F., \& {Asplund}, M.
  2009, \aap, 503, 545

\bibitem[{{Magain}(1984)}]{Magain1984}
{Magain}, P. 1984, \aap, 134, 189

\bibitem[{{Magrini} {et~al.}(2009){Magrini}, {Sestito}, {Randich}, \&
  {Galli}}]{Magrini2009}
{Magrini}, L., {Sestito}, P., {Randich}, S., \& {Galli}, D. 2009, \aap, 494, 95

\bibitem[{{Marino} {et~al.}(2015){Marino}, {Milone}, {Karakas}, {Casagrande},
  {Yong}, {Shingles}, {Da Costa}, {Norris}, {Stetson}, {Lind}, {Asplund},
  {Collet}, {Jerjen}, {Sbordone}, {Aparicio}, \& {Cassisi}}]{Marino2015}
{Marino}, A.~F., {Milone}, A.~P., {Karakas}, A.~I., {et~al.} 2015, \mnras, 450,
  815

\bibitem[{{Masseron} {et~al.}(2014){Masseron}, {Plez}, {Van Eck}, {Colin},
  {Daoutidis}, {Godefroid}, {Coheur}, {Bernath}, {Jorissen}, \&
  {Christlieb}}]{Masseron2014}
{Masseron}, T., {Plez}, B., {Van Eck}, S., {et~al.} 2014, \aap, 571, A47

\bibitem[{{Mateo} {et~al.}(1993){Mateo}, {Olszewski}, {Pryor}, {Welch}, \&
  {Fischer}}]{Mateo1993}
{Mateo}, M., {Olszewski}, E.~W., {Pryor}, C., {Welch}, D.~L., \& {Fischer}, P.
  1993, \aj, 105, 510

\bibitem[{{Mauro} {et~al.}(2012){Mauro}, {Moni Bidin}, {Cohen}, {Geisler},
  {Minniti}, {Catelan}, {Chen{\'e}}, \& {Villanova}}]{Mauro2012}
{Mauro}, F., {Moni Bidin}, C., {Cohen}, R., {et~al.} 2012, \apjl, 761, L29

\bibitem[{{McWilliam}(2016)}]{McWilliam2016}
{McWilliam}, A. 2016, \pasa, 33, e040

\bibitem[{{Mu{\~n}oz} {et~al.}(2017){Mu{\~n}oz}, {Villanova}, {Geisler},
  {Saviane}, {Dias}, {Cohen}, \& {Mauro}}]{Munoz2017}
{Mu{\~n}oz}, C., {Villanova}, S., {Geisler}, D., {et~al.} 2017, \aap, 605, A12

\bibitem[{{Mucciarelli} {et~al.}(2017){Mucciarelli}, {Monaco}, {Bonifacio}, \&
  {Saviane}}]{Mucciarelli2017}
{Mucciarelli}, A., {Monaco}, L., {Bonifacio}, P., \& {Saviane}, I. 2017, \aap,
  603, L7

\bibitem[{{Navarro} {et~al.}(2011){Navarro}, {Abadi}, {Venn}, {Freeman}, \&
  {Anguiano}}]{Navarro2011}
{Navarro}, J.~F., {Abadi}, M.~G., {Venn}, K.~A., {Freeman}, K.~C., \&
  {Anguiano}, B. 2011, \mnras, 412, 1203

\bibitem[{{Ness} {et~al.}(2013){Ness}, {Freeman}, {Athanassoula},
  {Wylie-de-Boer}, {Bland-Hawthorn}, {Asplund}, {Lewis}, {Yong}, {Lane}, \&
  {Kiss}}]{Ness2013}
{Ness}, M., {Freeman}, K., {Athanassoula}, E., {et~al.} 2013, \mnras, 430, 836

\bibitem[{{Nissen} {et~al.}(2014){Nissen}, {Chen}, {Carigi}, {Schuster}, \&
  {Zhao}}]{Nissen2014}
{Nissen}, P.~E., {Chen}, Y.~Q., {Carigi}, L., {Schuster}, W.~J., \& {Zhao}, G.
  2014, \aap, 568, A25

\bibitem[{{Nissen} {et~al.}(2000){Nissen}, {Chen}, {Schuster}, \&
  {Zhao}}]{Nissen2000}
{Nissen}, P.~E., {Chen}, Y.~Q., {Schuster}, W.~J., \& {Zhao}, G. 2000, \aap,
  353, 722

\bibitem[{{Nissen} \& {Schuster}(2010)}]{NissenSchuster2010}
{Nissen}, P.~E. \& {Schuster}, W.~J. 2010, \aap, 511, L10

\bibitem[{{Origlia} {et~al.}(2013){Origlia}, {Massari}, {Rich}, {Mucciarelli},
  {Ferraro}, {Dalessandro}, \& {Lanzoni}}]{Origlia2013}
{Origlia}, L., {Massari}, D., {Rich}, R.~M., {et~al.} 2013, \apjl, 779, L5

\bibitem[{{Pancino} {et~al.}(2010){Pancino}, {Carrera}, {Rossetti}, \&
  {Gallart}}]{Pancino2010}
{Pancino}, E., {Carrera}, R., {Rossetti}, E., \& {Gallart}, C. 2010, \aap, 511,
  A56

\bibitem[{{Pancino} {et~al.}(2017){Pancino}, {Romano}, {Tang}, {Tautvai{\v
  s}ien{\.e}}, {Casey}, {Gruyters}, {Geisler}, {San Roman}, {Randich},
  {Alfaro}, {Bragaglia}, {Flaccomio}, {Korn}, {Recio-Blanco}, {Smiljanic},
  {Carraro}, {Bayo}, {Costado}, {Damiani}, {Jofr{\'e}}, {Lardo}, {de Laverny},
  {Monaco}, {Morbidelli}, {Sbordone}, {Sousa}, \& {Villanova}}]{Pancino2017}
{Pancino}, E., {Romano}, D., {Tang}, B., {et~al.} 2017, \aap, 601, A112

\bibitem[{{Pehlivan Rhodin} {et~al.}(2017){Pehlivan Rhodin}, {Hartman},
  {Nilsson}, \& {J{\"o}nsson}}]{PehlivanRhodin2017}
{Pehlivan Rhodin}, A., {Hartman}, H., {Nilsson}, H., \& {J{\"o}nsson}, P. 2017,
  \aap, 598, A102

\bibitem[{{Placco} {et~al.}(2014){Placco}, {Frebel}, {Beers}, \&
  {Stancliffe}}]{Placco2014}
{Placco}, V.~M., {Frebel}, A., {Beers}, T.~C., \& {Stancliffe}, R.~J. 2014,
  \apj, 797, 21

\bibitem[{{Prusti} {et~al.}(2016){Prusti}, {de Bruijne}, {Brown}, {Vallenari},
  {Babusiaux}, {Bailer-Jones}, {Bastian}, {Biermann}, {Evans}, \&
  et~al.}]{GaiaDR1}
{Prusti}, T., {de Bruijne}, J.~H.~J., {Brown}, A.~G.~A., {et~al.} 2016, \aap,
  595, A1

\bibitem[{{Pryor} \& {Meylan}(1993)}]{Pryor1993}
{Pryor}, C. \& {Meylan}, G. 1993, in Astronomical Society of the Pacific
  Conference Series, Vol.~50, Structure and Dynamics of Globular Clusters, ed.
  S.~G. {Djorgovski} \& G.~{Meylan}, 357

\bibitem[{{Recio-Blanco} {et~al.}(2014){Recio-Blanco}, {de Laverny},
  {Kordopatis}, {Helmi}, {Hill}, {Gilmore}, {Wyse}, {Adibekyan}, {Randich},
  {Asplund}, {Feltzing}, {Jeffries}, {Micela}, {Vallenari}, {Alfaro}, {Allende
  Prieto}, {Bensby}, {Bragaglia}, {Flaccomio}, {Koposov}, {Korn}, {Lanzafame},
  {Pancino}, {Smiljanic}, {Jackson}, {Lewis}, {Magrini}, {Morbidelli},
  {Prisinzano}, {Sacco}, {Worley}, {Hourihane}, {Bergemann}, {Costado},
  {Heiter}, {Joffre}, {Lardo}, {Lind}, \& {Maiorca}}]{RecioBlanco2014}
{Recio-Blanco}, A., {de Laverny}, P., {Kordopatis}, G., {et~al.} 2014, \aap,
  567, A5

\bibitem[{{Reddy} {et~al.}(2016){Reddy}, {Lambert}, \& {Giridhar}}]{Reddy2016}
{Reddy}, A.~B.~S., {Lambert}, D.~L., \& {Giridhar}, S. 2016, \mnras, 463, 4366

\bibitem[{{Reddy} {et~al.}(2006){Reddy}, {Lambert}, \& {Allende
  Prieto}}]{Reddy2006}
{Reddy}, B.~E., {Lambert}, D.~L., \& {Allende Prieto}, C. 2006, \mnras, 367,
  1329

\bibitem[{{Reddy} {et~al.}(2003){Reddy}, {Tomkin}, {Lambert}, \& {Allende
  Prieto}}]{Reddy2003}
{Reddy}, B.~E., {Tomkin}, J., {Lambert}, D.~L., \& {Allende Prieto}, C. 2003,
  \mnras, 340, 304

\bibitem[{{Rojas-Arriagada} {et~al.}(2017){Rojas-Arriagada}, {Recio-Blanco},
  {de Laverny}, {Mikolaitis}, {Matteucci}, {Spitoni}, {Schultheis}, {Hayden},
  {Hill}, {Zoccali}, {Minniti}, {Gonzalez}, {Gilmore}, {Randich}, {Feltzing},
  {Alfaro}, {Babusiaux}, {Bensby}, {Bragaglia}, {Flaccomio}, {Koposov},
  {Pancino}, {Bayo}, {Carraro}, {Casey}, {Costado}, {Damiani}, {Donati},
  {Franciosini}, {Hourihane}, {Jofr{\'e}}, {Lardo}, {Lewis}, {Lind}, {Magrini},
  {Morbidelli}, {Sacco}, {Worley}, \& {Zaggia}}]{RojasArriagada2017}
{Rojas-Arriagada}, A., {Recio-Blanco}, A., {de Laverny}, P., {et~al.} 2017,
  \aap, 601, A140

\bibitem[{{Ruchti} {et~al.}(2016){Ruchti}, {Feltzing}, {Lind}, {Caffau},
  {Korn}, {Schnurr}, {Hansen}, {Koch}, {Sbordone}, \& {de Jong}}]{Ruchti2016}
{Ruchti}, G.~R., {Feltzing}, S., {Lind}, K., {et~al.} 2016, \mnras, 461, 2174

\bibitem[{{Schlafly} \& {Finkbeiner}(2011)}]{Schlafly2011}
{Schlafly}, E.~F. \& {Finkbeiner}, D.~P. 2011, \apj, 737, 103

\bibitem[{{Simmerer} {et~al.}(2004){Simmerer}, {Sneden}, {Cowan}, {Collier},
  {Woolf}, \& {Lawler}}]{Simmerer2004}
{Simmerer}, J., {Sneden}, C., {Cowan}, J.~J., {et~al.} 2004, \apj, 617, 1091

\bibitem[{{Simpson} {et~al.}(2017){Simpson}, {De Silva}, {Martell}, {Zucker},
  {Ferguson}, {Bernard}, {Irwin}, {Penarrubia}, \& {Tolstoy}}]{Simpson2017}
{Simpson}, J.~D., {De Silva}, G.~M., {Martell}, S.~L., {et~al.} 2017, \mnras,
  471, 4087

\bibitem[{{Sneden} {et~al.}(2008){Sneden}, {Cowan}, \& {Gallino}}]{Sneden2008}
{Sneden}, C., {Cowan}, J.~J., \& {Gallino}, R. 2008, \araa, 46, 241

\bibitem[{{Sneden}(1973)}]{Sneden1973}
{Sneden}, C.~A. 1973, PhD thesis, The University of Texas at Austin.

\bibitem[{{Spite} \& {Spite}(1982)}]{Spite1982}
{Spite}, F. \& {Spite}, M. 1982, \aap, 115, 357

\bibitem[{{Spitzer}(1987)}]{Spitzer1987}
{Spitzer}, L. 1987, {Dynamical evolution of globular clusters}

\bibitem[{{Van der Swaelmen} {et~al.}(2016){Van der Swaelmen}, {Barbuy},
  {Hill}, {Zoccali}, {Minniti}, {Ortolani}, \&
  {G{\'o}mez}}]{Vanderswaelmen2016}
{Van der Swaelmen}, M., {Barbuy}, B., {Hill}, V., {et~al.} 2016, \aap, 586, A1

\bibitem[{{Vande Putte} {et~al.}(2010){Vande Putte}, {Garnier}, {Ferreras},
  {Mignani}, \& {Cropper}}]{VandePutte2010}
{Vande Putte}, D., {Garnier}, T.~P., {Ferreras}, I., {Mignani}, R.~P., \&
  {Cropper}, M. 2010, \mnras, 407, 2109

\bibitem[{{Walker} {et~al.}(2007){Walker}, {Mateo}, {Olszewski}, {Gnedin},
  {Wang}, {Sen}, \& {Woodroofe}}]{Walker2007}
{Walker}, M.~G., {Mateo}, M., {Olszewski}, E.~W., {et~al.} 2007, \apjl, 667,
  L53

\bibitem[{{Woosley} \& {Weaver}(1995)}]{WoosleyWeaver1995}
{Woosley}, S.~E. \& {Weaver}, T.~A. 1995, \apjs, 101, 181

\bibitem[{{Wright} {et~al.}(2010){Wright}, {Eisenhardt}, {Mainzer}, {Ressler},
  {Cutri}, {Jarrett}, {Kirkpatrick}, {Padgett}, {McMillan}, {Skrutskie},
  {Stanford}, {Cohen}, {Walker}, {Mather}, {Leisawitz}, {Gautier}, {McLean},
  {Benford}, {Lonsdale}, {Blain}, {Mendez}, {Irace}, {Duval}, {Liu}, {Royer},
  {Heinrichsen}, {Howard}, {Shannon}, {Kendall}, {Walsh}, {Larsen}, {Cardon},
  {Schick}, {Schwalm}, {Abid}, {Fabinsky}, {Naes}, \& {Tsai}}]{Wright2010}
{Wright}, E.~L., {Eisenhardt}, P.~R.~M., {Mainzer}, A.~K., {et~al.} 2010, \aj,
  140, 1868

\bibitem[{{Wu} {et~al.}(2009){Wu}, {Zhou}, {Ma}, \& {Du}}]{Wu2009}
{Wu}, Z.-Y., {Zhou}, X., {Ma}, J., \& {Du}, C.-H. 2009, \mnras, 399, 2146

\bibitem[{{Zacharias} {et~al.}(2017){Zacharias}, {Finch}, \&
  {Frouard}}]{Zacharias2017}
{Zacharias}, N., {Finch}, C., \& {Frouard}, J. 2017, \aj, 153, 166

\end{thebibliography}
\end{document}